\documentclass[a4paper,11pt]{article}
\usepackage{jheppub} 
\usepackage{lineno}
\usepackage{hyperref}
\usepackage{graphicx} 
\usepackage{subcaption}
\usepackage{physics}
\usepackage{tikz}
\usepackage{quantikz}
\usepackage{diagbox}
\usepackage{booktabs}
\usepackage{multirow}
\usepackage{enumitem}

\title{A comparative study of variational quantum classifiers and classical machine learning for $tW$-$t\bar{t}$ discrimination at the LHC}

\author[a,1]{Rohan Hindocha\note{Corresponding author.}}
\author[a,2]{and Priyanka}
\affiliation[a]{Department of Physics, Hindu College, University of Delhi \\ Delhi 110007, India}

\emailAdd{rohanhindocha.464.2022@hinducollege.ac.in}

\abstract{Precise measurements of single top-quark production in the associated $tW$-channel require efficient discrimination against the dominant $t\bar{t}$ background. In this work, we investigate the applicability of quantum machine learning to this classification problem using the Variational Quantum Classifier (VQC) algorithm. Simulated proton-proton collision events from CMS Open Data corresponding to Run-2 are employed to train and evaluate the VQC model. Its performance is compared with two established classical approaches, Boosted Decision Trees (BDTs) and Multi-Layer Perceptron (MLP) using both machine learning metrics and physics-motivated metrics that describe the Signal Significance as a function of Signal Efficiency. The comparison is performed separately for the dileptonic $e-\mu$ channel and for the resolved and boosted hadronic topologies. In addition, signal-enriched and background-enriched phase-space regions are studied to assess the dependence of classifier performance on event topology. Our results provide a systematic benchmark for the application of VQC for a realistic High Energy Physics classification task and identify the analysis regimes wherein quantum-inspired models offer competitive discrimination with respect to established classical techniques. }


\begin{document}
\maketitle
\flushbottom

\section{Introduction}

The top quark occupies a unique position in the Standard Model owing to its large mass and corresponding strong Yukawa coupling to the Higgs Boson. As the heaviest known elementary particle, it provides a sensitive probe for Physics Beyond the Standard Model (BSM), including anomalous $Wtb$ coupling, Flavour Changing Neutral Currents and heavy resonances that preferentially couple to third-generation quarks \cite{49,50,52}. Consequently, precision measurements related to top quark production and decay remain one of the primary objectives of the Physics program at the Large Hadron Collider (LHC).

\vspace{3 pt}

Among the electroweak production mechanisms, the associated production of a single top quark with a W boson ($tW$) is of particular interest. The $tW$ process provides direct access to the $Wtb$ vertex and plays an important role in precision tests of the Standard Model, as well as in searches for new Physics \cite{49}. However, an accurate measurement of the $tW$ production cross-section remains experimentally challenging because its final-state topology closely resembles that of the top-antitop ($t\bar{t}$) production, which constitutes its dominant background \cite{49,50,52}.

\vspace{3 pt}

Machine-learning algorithms such as Boosted Decision Trees (BDTs) and Artificial Neural Networks (ANNs) have become standard tools in HEP analyses and have demonstrated excellent performance in event classification problems \cite{17,49,50,52}. More recently, advances in Quantum Computing have motivated the development of Quantum Machine Learning (QML) algorithms including Variational Quantum Classifier (VQC) algorithm, which employ Parameterized Quantum Circuits (PQCs) to construct non-linear decision boundaries in feature space \cite{32,33,35,36}. Although current quantum hardware remains limited, variational quantum algorithms have attracted considerable interest because of their compatibility with Noisy Intermediate-Scale Quantum (NISQ) devices and their potential applicability to realistic classification problems \cite{92}.

\vspace{3 pt}

The objective of this work is to investigate whether VQCs provide competitive performance for the $tW$ vs $t\bar{t}$ classification problem compared to established classical machine-learning algorithms. For this purpose, Monte-Carlo-simulated events from the CMS Open Data corresponding to Run-2 have been used. We compare the performance of VQCs with BDTs and Multi-Layer Perceptron (MLP) in three independent analysis topologies: the dileptonic $e-\mu$ channel, the resolved hadronic regime and the boosted hadronic regime. The comparison is performed using both conventional machine learning metrics (AUC and separation power) and Physics-motivated quantities, which bring out the characteristics of the curve that represents Signal Significance as a function of Signal Efficiency, thereby providing a systematic assessment of the applicability of VQC algorithm to a realistic HEP event-selection problem.

\section{\texorpdfstring{$tW$ vs  $t\bar{t}$ discrimination} {tW vs ttbar discrimination}}\label{section: tW vs tt}

\subsection{Top quark production at the LHC} \label{subsec: production}

\begin{figure}[h]
    \centering
    \includegraphics{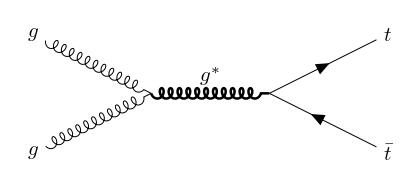}
    \caption{Feynman diagram of $t\bar{t}$ production}
    \label{fig: Figure 1}
\end{figure}

Top quarks are produced in proton-proton collisions at the Large Hadron Collider through two fundamental mechanisms: strong interaction and electroweak interaction. The dominant production mode is top-antitop ($t \bar{t}$) pair production, which proceeds through Quantum Chromodynamics (QCD). Primarily via gluon-gluon fusion,
$$gg \xrightarrow{} g^* \xrightarrow{} t \bar{t}$$ 
The corresponding leading-order Feynman diagram is shown in Figure \ref{fig: Figure 1} \cite{49,50,51,52,53}. 

\vspace{3 pt}

In addition to pair production, a single top quark can be produced through electroweak processes. At leading order, three production channels contribute:
$$q\bar{q}' \xrightarrow{} W^* \xrightarrow{} \bar{b}t $$
$$qb\xrightarrow{}q't$$
$$bg\xrightarrow{}tW$$
 \cite{49,50,51,52,53}

\begin{figure}[h]
\centering

\includegraphics{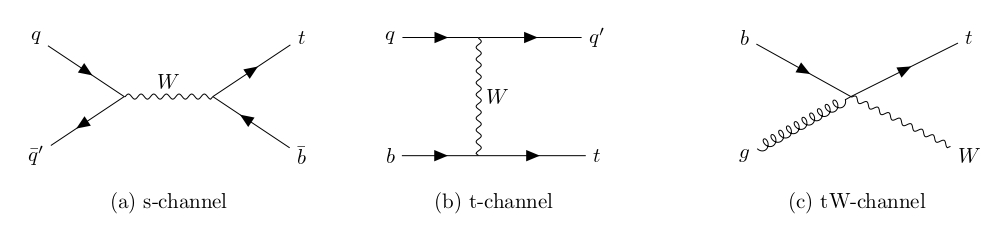}

\caption{Leading-order Feynman diagrams for single top-quark production: (a) s-channel, (b) t-channel, and (c) associated $tW$ production.}
\label{fig:Figure 2}

\end{figure}

The representative Feynman diagrams for these processes are presented in Figure \ref{fig:Figure 2}.
The production mechanisms are commonly classified according to the Mandelstam variables associated with the exchanged virtual particle. The process  $q\bar{q'} \xrightarrow{} \bar{b}t$ corresponds to the s-channel, where a virtual $W$ boson mediates the interaction between the incoming quark and anti-quark. The reaction $qb \xrightarrow{} q't$ represents the t-channel, in which the top quark is produced through the exchange of a virtual $W$ boson. In contrast, the associated production process $bg \xrightarrow{} tW$ is referred to as the tW channel and is not classified as one of the Mandelstam variables \cite{49,50,51,52,53}.

\vspace{3 pt}

\begin{table*}[h]
\centering
\resizebox{\textwidth}{!}{
\begin{tabular}{|c|cc|cc|cc|cc|}
\hline
 & \multicolumn{2}{c|}{7 TeV} & \multicolumn{2}{c|}{8 TeV} & \multicolumn{2}{c|}{13 TeV} & \multicolumn{2}{c}{13.6 TeV} \\
Process & Theory & Exp. & Theory & Exp. & Theory & Exp. & Theory & Exp. \\
\hline

$t\bar{t}$ 
& $\sim 177$ 
& $182.9 \pm 7.11$ 
& $\sim 242.9$ 
& $243.3 \pm 8.77$ 
& $\sim 832$ 
& $826.4 \pm 19.83$ 
& $\sim 924$ 
& $882 \pm 68$ \\

Single top (s) 
& $\sim 4.56$
& -- 
& $\sim 5.6$
& $4.8 \pm 0.8$ 
& $\sim 10.32$
& --
& $\sim 11.78$ 
& -- \\

Single top (t) 
& $\sim 64.6$ 
& $68 \pm 8.25$ 
& $\sim 87.8$ 
& $89.6 \pm 1.2$ 
& $\sim 217$ 
& $247 \pm 45.40$ 
& $\sim 232.2$
& -- \\

$tW$ 
& $\sim 15.7$ 
& $16.8 \pm 5.69$ 
& $\sim 22.4$ 
& $23 \pm 1.3$ 
& $\sim 71.7$ 
& $94 \pm 10$ 
& $\sim 87.9$ 
& $ 82.3 \pm 10.7$ \\

\hline
\end{tabular} }
\caption{Summary of top-quark production cross sections (in pb) at different center-of-mass energies. Experimental values include total uncertainties (statistical, systematic, and luminosity combined, uncertainties in theoretical values are not shown). (references for the data \cite{7,37,38,39,40,41,42,43,44,45,46,47,48})}
\label{tab:top_xsec}
\end{table*}

The production cross-sections for $t\bar{t}$ and single-top processes increase significantly with the center-of-mass energy of the collider. A comparison between theoretical predictions and experimental measurements at $\sqrt{s} = 7,8,13$ and $13.6 TeV$ is summarized in Table \ref{tab:top_xsec}. As expected, $t\bar{t}$ production exhibits the largest cross-section, whereas electroweak single-top channels occur less frequently. 

Due to its large mass, the top quark ($172.60 \pm 0.27 GeV/c^{2}$ \cite{PDG2026}) decays before hadronization. Its lifetime ( $10^{-25}$ s) is shorter than the typical hadronization timescale ($10^{-24}$ s), allowing the properties of a bare quark to be studied directly. The dominant decay mode is $t\xrightarrow{} Wb$ \cite{49,50,51,52} . Consequently, the complete decay chain becomes
$$t\bar{t} \xrightarrow{} WbW\bar{b}$$
for top-pair production and
$$tW \xrightarrow{} WbW$$
for associated single-top production \cite{49,50,51,52,53}. 

\vspace{3 pt}

Although these final states appear distinct at parton level, detector level reconstruction introduces important complications. Initial-State Radiation (ISR) and Final-State Radiation (FSR) can produce additional b-tagged jets in $tW$ events. Conversely, in $t\bar{t}$ events, one of the b-jets may escape reconstruction because of low transverse momentum or detector acceptance, causing the event to resemble a $tW$ topology \cite{17,51,52}. As a result, both processes can populate nearly identical reconstructed final states, making their separation a challenging classification problem that requires the use of discriminating kinematic observables and advanced Machine-Learning techniques.

\subsection{Decay modes}\label{subsec:decay modes}
\begin{figure}[h]
\centering
\includegraphics{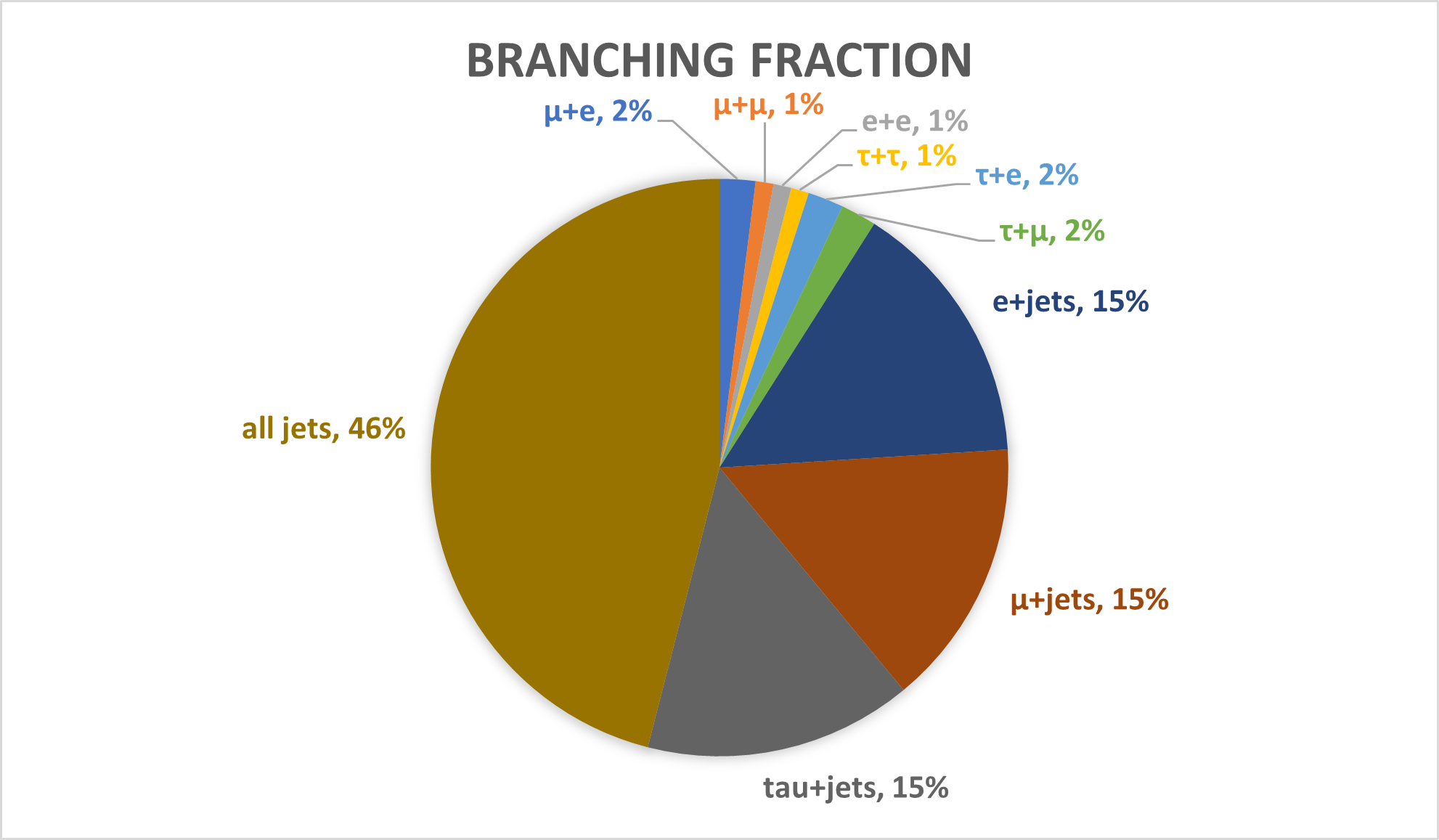}
\caption{Branching fractions of $tW$ decay modes.\label{fig:tW decay modes}}
\end{figure}
\vspace{3 pt}
Following the production of a top quark in the associated tW process, the top quark predominantly decays into a $W$ boson and a bottom quark, while the accompanying $W$ boson subsequently undergoes either a leptonic or a hadronic decay. Consequently, the complete event topology is determined by the decay modes of both W bosons, giving rise to several experimentally distinguishable final states \cite{49,50,51,52,53}.

\vspace{3 pt}

The relative branching fractions of these decay channels are illustrated in Figure \ref{fig:tW decay modes}. Depending on the decay products of the two $W$ bosons, the final states can be broadly categorized into  fully leptonic, semi-leptonic, and fully hadronic topologies.

\vspace{3 pt}

In this work, we focus on two representative channels that are widely used in experimental top-quark analyses. 
$$W \xrightarrow{} e  + \nu_{e} $$
$$W \xrightarrow{} \mu + \nu_{\mu}$$
The first is the dileptonic $e-\mu$ channel, in which the two $W$ bosons decay into an electron and a muon (together with the corresponding neutrinos). This channel offers a relatively clean experimental signature and benefits from a reduced contribution from multi-jet background processes.

\vspace{3 pt}

The second is the fully hadronic (all-jets) channel, where both $W$ bosons decay into quark pairs.
$$W \xrightarrow{} q \bar{q'}$$
Although this decay mode has the largest branching fraction, it is experimentally more challenging because of the large background arising from QCD multi-jet production. Nevertheless, it provides valuable information because of its high event yield and is therefore included in the present analysis.

\vspace{3 pt}

The classification performance of Machine-Learning models is evaluated independently for these two decay channels to investigate their topologies and background conditions.

\subsection{Regions for analysis} \label{subsec: regions}

The present analysis is performed in two independent final states: the dileptonic $e-\mu$ channel and the fully hadronic channel. These channels exhibit distinct event topologies and therefore require different event categorization strategies.

\subsubsection{\texorpdfstring{ Dileptonic $e-\mu$ channel}{e-mu channel}}
In the dileptonic channel, both $W$ bosons decay leptonically, $ W \xrightarrow{} l \nu$ leading to the signal process 
$$ tW \xrightarrow{} WbW \xrightarrow{} l \nu l \nu b$$,
whereas the dominant background proceeds through 
$$ t\bar{t} \xrightarrow{} WbW\bar{b} \xrightarrow{} l \nu l \nu b\bar{b}$$.
At leading order, the final state is characterized by a single b-jet whereas the dominant $t\bar{t}$ background yields two b-jets. However, additional jets originating from ISR and FSR, together with detector inefficiencies in jet reconstruction and b-jet tagging, often modify the observed jet multiplicity. Consequently, the analysis phase space is divided into three mutually exclusive event categories:
\begin{enumerate}
\item{1j1b: exactly one reconstructed jet, which is b-tagged (signal-enriched region)}
\item{2j1b: exactly two reconstructed jets with one b-tagged jet (mixed region)}
\item{2j2b: exactly two reconstructed b-tagged jets (background).}
\end{enumerate}  \cite{7,17,42}

The 2j1b region contains the largest fraction of reconstructed events and therefore serves as the primary signal sensitive region, whereas the 2j2b category is dominated by $t\bar{t}$ production and provides a valuable control region to evaluate classifier performance \cite{17}.

\subsubsection{Hadronic channel}
For the fully hadronic channel, both $W$ bosons decay into quarks ($W \xrightarrow{} qq'$), resulting in 
$$tW \xrightarrow{}bqq'q''q'''$$ and 
$$t\bar{t} \xrightarrow{} b\bar{b}qq'q''q'''$$.
 
The event topology depends strongly on the transverse momentum of the top quark. At relatively low transverse momentum, the decay products are well separated and reconstructed as individual AK4 jets (R=0.4). This topology is referred to as resolved regime. In contrast, highly boosted top quarks produce collimated decay products that merge into a single large-radius AK8 jet (R=0.8) , defining the boosted regime \cite{82,83,84}. 

Within both regimes, events are further classified according to the number of reconstructed b-tagged jets:
\begin{enumerate}
\item{1b: one reconstructed b-tagged jet (signal-enriched region)}
\item{2b: two or more reconstructed b-tagged jets (background-enriched region)}.
\end{enumerate}

This categorization enables Machine-Learning algorithms to exploit differences in event topology while simultaneously studying their performance in signal- and background-dominated regions \cite{82,83,84}.

\subsection{\texorpdfstring{Phase-space overlap and treatment of $tW- t\bar{t}$ interference}{phase-space overlap}}
 \begin{figure}[h]
\centering
\includegraphics{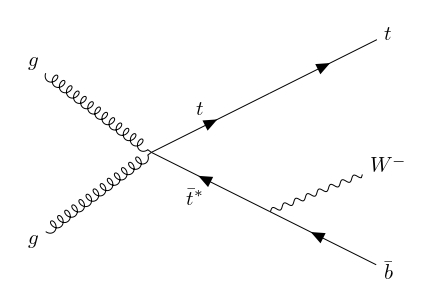}
\caption{Next-to-Leading-Order Feynman diagram of tW-channel top production}
\label{fig:NLO_tW_feynman}
\end{figure}
At leading order, the $tW$ and $t\bar{t}$ production mechanisms are described by distinct Feynman diagrams and correspond to different physical processes (refer to figures \ref{fig: Figure 1},\ref{fig:Figure 2} and subsection \ref{subsec: production}). However, next-to-leading order (NLO), additional diagrams contribute to the $tW$ production process $gg\xrightarrow{}tW\bar{b}$, some of which contain an intermediate on-shell top quark satisfying $m_{Wb} \approx m_t$. These resonant diagrams become indistinguishable from the corresponding $t\bar{t}$ production diagrams, $gg\xrightarrow{}t\bar{t}\xrightarrow{}tW\bar{b}$, leading to an overlap between the two Monte-Carlo samples (refer to figure \ref{fig:NLO_tW_feynman}) . If untreated, the same physical events would be counted twice, resulting in an incorrect prediction of both signal yield and kinematic distributions \cite{17,18,42}.

\vspace{3 pt}

Two standard approaches are commonly adopted to remove this ambiguity: the Diagram Removal (DR) scheme and the Diagram Subtraction (DS) scheme. In the DR approach, resonant $t\bar{t}$-like diagrams are removed directly from the $tW$ matrix element, whereas the DS method introduces a local subtraction term that preserves interference effects while avoiding double counting\cite{17,18,42}. 

\vspace{3 pt}

Throughout the work, the Diagram Removal (DR) scheme is employed for the simulated $tW$ samples, following the standard CMS event production procedure. This treatment provides a well-defined separation between the $tW$ signal and the dominant $t\bar{t}$ background used to train and evaluate Machine-Learning classifiers.

\section{Variational Quantum Classifier}\label{sec:qml}
A Variational Quantum Classifier is a hybrid quantum-classical learning framework in which classical input variables are encoded into a parameterized quantum circuit. The circuit transforms the encoded quantum state through trainable operations, and measurements of the final state provide quantities that can be used for classification \cite{28,30,31,32,33}. 

In the present study, the VQC is used as a binary classifier to distinguish $tW$ signal events from the $t\bar{t}$ background. The trainable parameters of the circuit are optimized using the corresponding event labels to minimize the classification-loss. The overall procedure therefore combines quantum-state evolution with classical parameter optimization.


\subsection{Theory} \label{subsec:qmltheory}
Classical Neural Networks optimize trainable weights and biases via gradient-based methods to minimize a loss function over iterations. Information is represented classically as bits that propagate through successive layers.

\vspace{3 pt}
Quantum Neural Networks encode information into qubits described by superposition of basis states $\ket{1}$ and$\ket{0}$. 
$\ket{\psi} = \alpha\ket{0}+\beta\ket{1}$
for 2 qubits the computational basis expands to $\ket{00} , \ket{01}, \ket{10}, \ket{11}$ and similarly for n qubits, the Hilbert space dimension grows to $2^n$ basis kets. The information is encoded as the superposition of $2^n$ basis vectors, each with some amplitudes whose squares add up to 1. Now, qubits are entangled in a desired manner in order to learn complex correlations. The angles on the Bloch sphere become our trainable parameters. These parameters control the amplitudes of each basis vector. The amplitudes are used to find the expectation value of the complete state of the system, which is used to compute the loss function. The optimization algorithms update the parameters according to the value of the loss function before each iteration, and finally the classification is based on minimization of the loss function\cite{22,24,26,28,30,31,32,54}.

\vspace{3 pt}
This architecture has special entanglement feature that can lead to learning of complex correlations between features, like non-factorizable relations which might offer the quantum advantage over classical neural networks.\cite{33,34,35,36} 
\subsection{Quantum Circuits}
\subsubsection{Rotation and Encoding}
For n qubits, the Hilbert space is represented as \[
\mathcal{H} = (\mathbb{C}^2)^{\otimes n}
\]
The basis vector spans the Hilbert space, and the general state is written as follows:
\[
|\Psi\rangle = \sum_{i=0}^{2^n-1} c_i |i\rangle, \quad \sum_i |c_i|^2 = 1
\]
where $c_i$ are the amplitudes for each basis vector.
 $\theta$ serves as the trainable parameter and the information is encoded in qubits through the change of $theta$. This requires rotation gates. Thus, Encoding operator is given as follows:
\[
U_{\text{enc}}(x) = \bigotimes_{k=1}^n R_y(x_k)
\]
The wavefunction encoded is given as follows:
\[
|\Psi_{\text{enc}}\rangle = U_{\text{enc}}(x)\,|0\rangle^{\otimes n}
\]
We note that the rotation operator is 
\[
R_y(x_k) = e^{-i x_k Y /2}
\]
where
\[
Y =
\begin{pmatrix}
0 & -i \\
i & 0
\end{pmatrix}
\]
thus
\[
R_y(\theta) =
\begin{pmatrix}
\cos(\theta/2) & -\sin(\theta/2) \\
\sin(\theta/2) & \cos(\theta/2)
\end{pmatrix}
\]
 \[
|\Psi\rangle =
c_{00}|00\rangle + c_{01}|01\rangle + c_{10}|10\rangle + c_{11}|11\rangle
\] \cite{22,28,32}
\begin{figure}[h]
\centering
\includegraphics[width=0.7\textwidth]{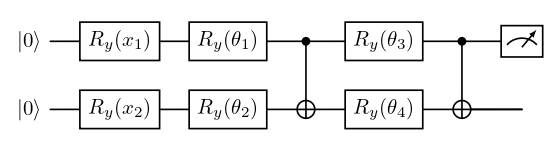}
\caption{Two Qubit example of Parameterized Quantum Circuit.}
\label{figure:quantum_circuit}
\end{figure}
\subsubsection{Entanglement}
After the rotations, the states are not entangled yet. The encoded wavefunction can be expressed as a tensor product of individual qubit wavefunctions.
\[
|\Psi_{\text{enc}}\rangle = \bigotimes_{k=1}^n |\psi_k\rangle
\]
Quantum Entanglement occurs when the encoded state can no longer be expressed as a tensor product of the individual qubit wavefunctions.
\[
|\Psi_{\text{enc}}\rangle \neq \bigotimes_{k=1}^n |\psi_k\rangle
\]
For this the CNOT gate is used.
\[
U_{\text{CNOT}} =
\begin{pmatrix}
1 & 0 & 0 & 0 \\
0 & 1 & 0 & 0 \\
0 & 0 & 0 & 1 \\
0 & 0 & 1 & 0
\end{pmatrix}
\]
Quantum entanglement property is exploited to allow the qubits to learn complex correlations. After this successive layers of rotation and entanglement are used for the learning process, the final state is given by
\[
|\Psi(\theta, x)\rangle =
U(\theta)\,U_{\text{enc}}(x)\,|0\rangle
\]
where 
\[
U(\theta) = \prod_{\ell=1}^L U_{\text{ent}}^{(\ell)} \, U_{\text{rot}}^{(\ell)}(\theta_\ell)
\]
\[
U_{\text{rot}}^{(\ell)} =
\bigotimes_{k=1}^n R_y(\theta_k^{(\ell)}) R_z(\theta_k'^{(\ell)})
\]
is the rotating layer operator and
\[
U_{\text{ent}}^{(\ell)} = \prod_{(i,j)\in E} \text{CNOT}_{ij}
\]
is the entangling layer operator \cite{23,32}.
\subsubsection{Measurement}
Then the function to be estimated is given by 
\[
f_\theta(x) =
\langle \Psi(\theta,x) | O | \Psi(\theta,x)\rangle
\]
where
\[
O = I \otimes \cdots \otimes Z_k \otimes \cdots \otimes I
\]
$Z$ is the Pauli operator which performs the measurement operation 
\[
Z =
\begin{pmatrix}
1 & 0 \\
0 & -1
\end{pmatrix}
\]\cite{32}
We obtain a set of n values: $$\{z_1, z_2,...z_n\}$$ where $$z_i= \bra{\psi(x,\theta)}Z_i \ket{\psi(x,\theta)}$$ gives the expectation value of the Pauli measurement operator for the $i^{th}$ qubit. The z value completely depends upon the state vector and lies in the range $[-1,+1]$.
Let us take an example from Figure \ref{figure:quantum_circuit} to understand this. 
\begin{enumerate}
    \item{Data is encoded by applying rotation gates on two qubits.}
    \item{Another layer of rotation gates is used to introduce parameter $\theta_1$ and $\theta_2$.}
    \item{CNOT gate is used to entangle the two qubits.}
    \item{Another set of parameters $\theta_3$ and $\theta_4$ is introduced. This time the system is in entangled state and rotation operation on one qubit directly alters the amplitudes in the other.}
    \item{Another layer of entanglement again allows for exchange of new information among qubits}
    \item{After these we perform measurement to estimate the function \cite{25,54}.}
\end{enumerate} 
\subsection{Optimizer}
After measurement, the optimizer maps the estimated function with the labels provided for classification and calculates the parameters to be updated in next iteration so as to minimize the loss function and bring the estimated function closer to the set of labels.\cite{21,28,30,33}

\vspace{3 pt}
Employing too many qubits or excessive entanglement layers can drive the VQC into a barren plateau where the gradient of loss function is very small. Hence, parameters are not correctly updated and the system despite of many iterations is stuck with the same final state. This occurs due to over-entanglement. A lot of information being exchanged renders the state essentially random hence no new learning takes place. \cite{27,29}
\begin{figure}[h]
\centering
\includegraphics[width=0.7\textwidth]{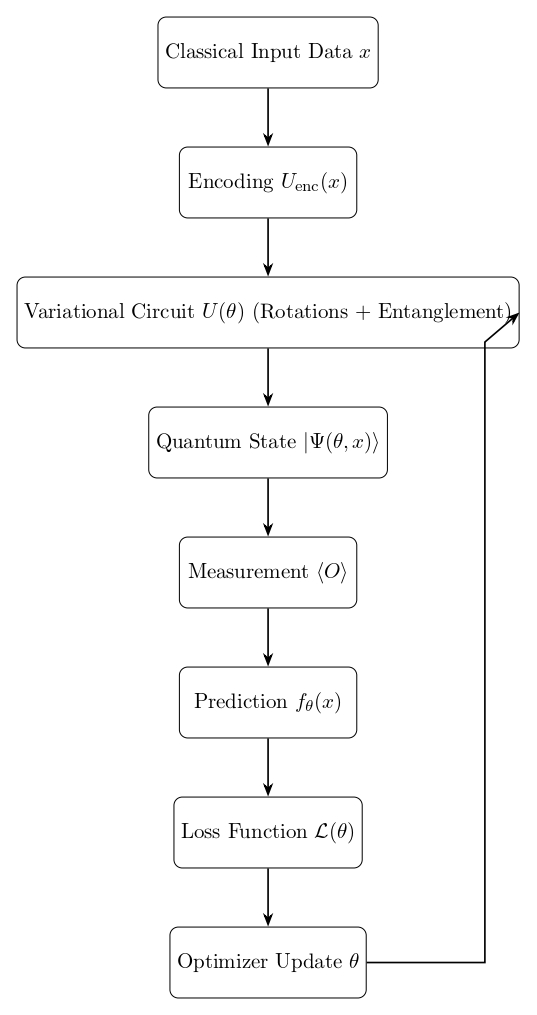}
\caption{Working of Parameterized Quantum Circuit.}
\label{figure:block_diagram}
\end{figure}
\subsection{Scaling of Data}
Since rotation gates are periodic operations, unscaled features with large dynamic ranges can lead to equivalent angles mapping to vastly different physical states, thereby degrading learning stability.

\[
R_y(x_k) = e^{-i x_k Y /2}
\]
We, therefore scale all features to the range $[0,\pi]$.
\[
x_{\text{scaled}} = \pi \cdot \frac{x - x_{\min}}{x_{\max} - x_{\min}}
\]
\subsection{Simulation}
For our paper, we use the \emph{default.qubit} PennyLane simulator for the simulation of VQC with PyTorch interface. It is an ideal state vector simulator and models the state vectors and their evolution with vector algebra. As an ideal state vector simulator, it does not account for gate noise, decoherence or readout errors present on  NISQ hardware. Consequently, our results represent an upper bound on VQC performance. Full setup of VQC is discussed in the section \ref{subsec: vqc methodology}. The PyTorch interface helps for mathematical operations such as gradient calculations during training and keeps the datatype compatible with the optimizer input for smooth transfer of data from quantum circuit simulator to optimizer for loss function calculation and parameter updates. We use the Adam optimizer in our study.

The simulator simulates quantum circuits with corresponding vector algebra operations. For example, one of the basis vectors $\ket{\psi}$ can be simulated as 
\[\ket{\psi} = 
\begin{pmatrix}
    1\\
    0
\end{pmatrix}\] for a system with 2 qubits. If this basis vector is to be operated with rotation $R_y(\theta)$, the simulator carries out matrix multiplication to give $R_y(\theta)\ket{\psi}$.
For $\theta = \pi/2$ and $\ket{\psi}$ as basis vector described above this becomes 
\[
R_y(\theta)\ket{\psi} =
\begin{pmatrix}
\cos(\pi/4) & -\sin(\pi/4) \\
\sin(\pi/4) & \cos(\pi/4)
\end{pmatrix}
\begin{pmatrix}
1\\
0
\end{pmatrix}=\frac{1}{\sqrt{2}}
\begin{pmatrix}
1\\
1
\end{pmatrix}
\]
Similarly, all other quantum circuit logic are carried out via linear algebra operators.

\section{Methodology}
\subsection{Datasets}\label{subsec:datasets}
Table \ref{tab:datasets} shows all the datasets used for analysis. The MC samples are taken from CERN open data portal CMS-Run2 campaign 2016 (13 TeV). For each process, a representative MINIAODSIM file from the CMS Open Data portal is utilized. While this provides a statistically limited sample, it is sufficient for a proof of concept comparison of classifier architectures.

\vspace{3 pt}

The plots in Figure \ref{fig:e-mu1} and \ref{fig:e-mu2} are plotted using the same datasets. For the $e-\mu$ channel single top and anti-top $tW$ datasets form the signal while $t\bar{t}$ and $W+jets$ form the background. Figure \ref{fig:e-mu1} shows the suppression of $W+jets$ which clearly has a large cross-section as compared to $tW$ or $t\bar{t}$ (refer Table \ref{tab:datasets} for cross-sections of each event). 

\vspace{3 pt}
For the hadronic channel, $tW$ top and anti-top form the signal while $t\bar{t}$, $W+jets$,$t-channel$ top and anti-top and all datasets of QCD jets form the backgrounds. Diagram Removal (DR) scheme is applied to $tW$ samples for simulation.While $tW$, $t\bar{t}$ and $t-channel$ samples are generated with POWHEG at NLO, $W+jets$, QCD jets and $t\bar{t}+jets$ samples are generated using MadGraphMLM at LO. PYTHIA8 is used to for parton showering and hadronization. 

\vspace{3 pt}
The study focuses only on background discrimination using ML techniques for the $t\bar{t}$ background while plots in Figure \ref{fig:e-mu1} and \ref{fig:e-mu2} summarize the suppression of rest of the backgrounds of $tW$ with Physics cuts.
\begin{table}[h!]
\centering
\hspace*{-2cm}
\begin{tabular}{|c|c|c|c|}
\hline
\textbf{Process} & \textbf{Sample} & \textbf{Generator} & \textbf{Cross-section (pb)} \\
\hline

\multicolumn{4}{|c|}{\textbf{e-$\mu$ Channel}} \\
\hline

$t\bar{t}$ & TTTo2L2Nu ~\cite{62} & POWHEG + PYTHIA8 & $6.871\times10^{2}$ \\
$tW$ top & ST\_tW\_top ~\cite{63} & POWHEG + PYTHIA8 & $3.245\times10^{1}$ \\
$tW$ antitop & ST\_tW\_antitop ~\cite{64} & POWHEG + PYTHIA8 & $3.251\times10^{1}$ \\
$W+$jets & WJetsToLNu ~\cite{65} & MadGraphMLM + PYTHIA8 & $5.411\times10^{4}$ \\
$WW$ & WWTo2L2Nu ~\cite{66} & POWHEG + PYTHIA8 & $1.109\times10^{1}$ \\
\hline

\multicolumn{4}{|c|}{\textbf{Hadronic Channel}} \\
\hline

$tW$ top & ST\_tW\_top ~\cite{67} & POWHEG + PYTHIA8 & $3.245\times10^{1}$ \\
$tW$ antitop & ST\_tW\_antitop ~\cite{68} & POWHEG + PYTHIA8 & $3.251\times10^{1}$ \\
$t\bar{t}$ & TTToHadronic ~\cite{69} & POWHEG + PYTHIA8 & $6.871\times10^{2}$ \\
$t\bar{t}+$jets & TTJets ~\cite{70} & MadGraphMLM + PYTHIA8 & $4.755\times10^{2}$ \\
$W+$jets & WJetsToLNu ~\cite{71} & MadGraphMLM + PYTHIA8 & $5.411\times10^{4}$ \\
$t$-channel top & ST\_t-channel\_top ~\cite{72} & POWHEG + PYTHIA8 & $1.197\times10^{2}$ \\
$t$-channel antitop & ST\_t-channel\_antitop ~\cite{73} & POWHEG + PYTHIA8 & $7.174\times10^{1}$ \\
\hline

\multicolumn{4}{|c|}{\textbf{QCD Multijet Background}} \\
\hline

QCD HT100--200 & QCD\_HT100to200 ~\cite{74} & MadGraphMLM + PYTHIA8 & $2.362\times10^{7}$ \\
QCD HT200--300 & QCD\_HT200to300 ~\cite{75} & MadGraphMLM + PYTHIA8 & $1.553\times10^{6}$ \\
QCD HT300--500 & QCD\_HT300to500 ~\cite{76} & MadGraphMLM + PYTHIA8 & $3.247\times10^{5}$ \\
QCD HT500--700 & QCD\_HT500to700 ~\cite{77} & MadGraphMLM + PYTHIA8 & $3.035\times10^{4}$ \\
QCD HT700--1000 & QCD\_HT700to1000 ~\cite{78} & MadGraphMLM + PYTHIA8 & $6.425\times10^{3}$ \\
QCD HT1000--1500 & QCD\_HT1000to1500 ~\cite{79} & MadGraphMLM + PYTHIA8 & $1.117\times10^{3}$ \\
QCD HT1500--2000 & QCD\_HT1500to2000 ~\cite{80} & MadGraphMLM + PYTHIA8 & $1.083\times10^{2}$ \\
QCD HT2000--Inf & QCD\_HT2000toInf ~\cite{81} & MadGraphMLM + PYTHIA8 & $2.194\times10^{1}$ \\
\hline

\end{tabular}
\caption{Monte Carlo simulated datasets used in the analysis obtained from the CMS Open Data portal.}
\label{tab:datasets}
\end{table}

\subsection{Weights and Statistics before training} \label{subsec: stats}

\begin{table}[h!]
\centering
\resizebox{\textwidth}{!}{
\begin{tabular}{|l|c|c|c|c|c|c|}
\hline
\textbf{Event Category} 
& \multicolumn{2}{c|}{\textbf{tW top}} 
& \multicolumn{2}{c|}{\textbf{tW antitop}} 
& \multicolumn{2}{c}{\textbf{$t\bar{t}$}} \\
\cline{2-7}
& \textbf{Events} & \textbf{Weighted Yield ($10^6$)} 
& \textbf{Events} & \textbf{Weighted Yield($10^6$)} 
& \textbf{Events} & \textbf{Weighted Yield($10^6$)} \\
\hline
Non-hadronic (MINIAODSIM) & 80713  & 2.619  & 90817 & 2.952 & 46000 & 3.316  \\
e-$\mu$ channel        & 5851  & 0.085   & 6750 & 0.087 & 8907 & 4.791 \\
Hadronic (MINIAODSIM)     & 53000  & 1.720 & 29000 & 0.943  & 17000 & 5.342 \\
Boosted region         & 7154   & 0.158  & 3952  & 0.159  & 3113  &  4.506  \\
Resolved region        & 27289  & 0.601  & 14983 & 0.605  & 12443 & 18.130 \\
\hline
\end{tabular}}
\caption{Event counts and weighted yields used for training in different channels}
\label{table:event_counts}
\end{table}

\begin{table}[h!]
\centering
\begin{tabular}{|l|c|c|c|c|}
\hline
\textbf{Region}  &  \multicolumn{2}{c|}{\textbf{Signal}} & \multicolumn{2}{c|}{\textbf{Background}}\\
\cline{2-5}
&\textbf{Events} & \textbf{Weighted Yield} 
&\textbf{Events} & \textbf{Weighted Yield}\\
\hline
1j1b & 24   &  322 & 10 & 5423   \\
2j1b & 6475 & 88230 & 4374 & 2355874 \\
2j2b & 1738  & 23632 & 2694 & 1446932 \\
\hline
\end{tabular}
\caption{Signal and background events and weighted yield in different regions of $e-\mu$ channel}
\label{table:region_events_leptonic}
\end{table}

\begin{table}[h!]
\centering
\begin{tabular}{|l|c|c|c|c|}
\hline
\textbf{Region} & \multicolumn{2}{c|}{\textbf{Signal}} & \multicolumn{2}{c}{\textbf{Background}} \\
\cline{2-5}
 & \textbf{Events} & \textbf{Weighted Yield} & \textbf{Events} & \textbf{Weighted Yield} \\
\hline
1b & 34283  & 977489  & 7516  & 10949312  \\
2b & 7989 & 228675  & 4927 & 7178892 \\
\hline
\end{tabular}
\caption{Signal and background events in different regions of hadronic resolved channel}
\label{table:region_events_hadronic_resolved}
\end{table}

\begin{table}[h!]
\centering
\begin{tabular}{|l|c|c|c|c|}
\hline
\textbf{Region} & \multicolumn{2}{c|}{\textbf{Signal}} & \multicolumn{2}{c}{\textbf{Background}} \\
\cline{2-5}
 & \textbf{Events} & \textbf{Weighted Yield} & \textbf{Events} & \textbf{Weighted Yield} \\
\hline
1b & 5497  & 157425  & 1416  & 2048599  \\
2b & 2573 & 73419  & 1126 & 1632432 \\
\hline
\end{tabular}
\caption{Signal and background events in different regions of hadronic boosted channel}
\label{table:region_events_hadronic_boosted}
\end{table}
Table \ref{table:event_counts} conveys the total events in hadronic and non-hadronic MINIAODSIM root files and the subsets of events that fall into desired channels of analysis. Table \ref{table:region_events_leptonic} , Table \ref{table:region_events_hadronic_resolved} and Table \ref{table:region_events_hadronic_boosted} contain information about events in each region of analysis in the $e-\mu$, hadronic resolved and hadronic boosted channels respectively. Note from Table \ref{table:region_events_leptonic} region 1j1b in $e-\mu$ channel does not contain sufficient number (24) of events and therefore the analysis in this event is not considered. The cuts placed to enforce these channels are:

\textbf{$e-\mu$ channel:}
\begin{itemize}[itemsep=1.5pt, topsep=3pt]

\item At least one muon satisfying
\[
p_T^\mu > 10~\mathrm{GeV}
\]

\item At least one electron satisfying
\[
p_T^e > 10~\mathrm{GeV}
\]

\item Require at least one electron and one muon:
\[
N_\mu \geq 1,\qquad N_e \geq 1
\]

\item Opposite-sign dilepton requirement:
\[
q_\mu \times q_e < 0
\]

 \item Leading lepton transverse momentum:
  \[
 p_T^{\ell_1} > 25~\mathrm{GeV}.
  \]

 \item Subleading lepton transverse momentum:
 \[
 p_T^{\ell_2} > 20~\mathrm{GeV}.
 \]

\item Lepton pseudorapidity:
 \[
 |\eta_\ell| < 2.5.
 \]

 \item Missing transverse energy:
 \[
 E_T^{\rm miss} > 30~\mathrm{GeV}.
 \]

\item Angular separation between the electron and muon:
\[\Delta R(e,\mu) > 0.3 \]
   
\item AK4 jet selection:
\[
p_T^{\mathrm{jet}} > 30~\mathrm{GeV}
\]

\item b-jet definition using CSVv2 discriminator:
\[
\mathrm{CSVv2} > 0.8
\]

\end{itemize}

The cuts placed for selection of events for the $e-\mu$ topology are motivated by CMS analyses papers \cite{6,7,93}. The numerical selection thresholds were modified where necessary to retain sufficient event statistics while preserving the characteristic topology.

\textbf{Hadronic Boosted channel:}
\begin{itemize}[itemsep=1.5pt, topsep=3pt]

\item Require at least one AK8 fatjet satisfying
\[
p_T^{\mathrm{AK8}} > 200~\mathrm{GeV}
\]
and
\[
|\eta^{\mathrm{AK8}}| < 2.4
\]

\item Event rejected unless AK8 jet contains valid substructure observables:
\[
\tau_1 (1-subjettiness),\qquad \tau_2(2-subjettiness),\qquad m_{\mathrm{SD}}(softdrop mass)
\]

\item AK4 jet selection:
\[
p_T^{\mathrm{jet}} > 30~\mathrm{GeV}
\]

\item Require at least two AK4 jets:
\[
N_{\mathrm{jets}} \geq 2
\]

\item b-jet definition using CSVv2 discriminator:
\[
\mathrm{CSVv2} > 0.8
\]

\end{itemize}

The cuts placed for selection of events for the hadronic boosted topology are motivated by CMS analyses papers \cite{83,84,94,95}. The numerical selection thresholds were modified where necessary to retain sufficient event statistics while preserving the characteristic topology.

\textbf{Hadronic Resolved channel:}
\begin{itemize}[itemsep=1.5pt, topsep=3pt]

 \item At least four AK4 jets.

 \item Jet transverse momentum:
 \[
 p_T^{\mathrm{jet}} > 30~\mathrm{GeV}.
 \]
 \item Jet pseudorapidity:
 \[
 |\eta_{\mathrm{jet}}| < 2.5.
  \]
    
\end{itemize}

The cuts placed for selection of events for the hadronic resolved topology are motivated by CMS analysis paper \cite{96}.

\subsection{Feature-set for training} \label{subsec: features}
The features used for training the three ML models: BDT, MLP and VQC are discussed in this section. Tables \ref{tab:separation_power e-mu}, \ref{tab:separation_power hadronic resolved}, \ref{tab:separation_pwer hadronic boosted} display the feature sets with their separating power in different regions for $e-\mu$ channel, hadronic resolved channel and hadronic boosted channel respectively. As discussed in subsection \ref{subsec: regions}, the phase space has been divided into regions 1j1b, 2j1b and 2j2b in the $e-\mu$ channel, while in the hadronic channels regions 1b of 2b of the phase space are used for the analysis. Due to insufficient statistics in region 1j1b (refer Table \ref{table:region_events_leptonic}), we drop it from the analysis. Region 2j1b serves as the signal-dominated region while 2j2b as background-dominated region in the $e-\mu$ channel. Similarly, region 1b acts as signal-dominated region  and 2b as background-dominated in hadronic channels. Since we have selected events with 4 or more jets for the hadronic resolved regime, it was possible to have the transverse momentum of the 3rd jet as a training feature.

\emph{Separation Power} quantifies how well a variable can be distinguished between two classes, such as signal and background.
\[
\mathcal{S}
=
\frac{1}{2}
\int
\frac{\left[S(x)-B(x)\right]^2}
{S(x)+B(x)}
\,dx
\]

\begin{table}[h!]
\centering
\begin{tabular}{|c|c|c|}
\hline
\textbf{Feature} & \multicolumn{2}{c|}{\textbf{Separation Power}} \\ 
\cline{2-3}
 & \textbf{2j1b} & \textbf{2j2b} \\ 
\hline
ST  & 0.0039 & 0.0436 \\ \hline
HT  & 0.0098 & 0.0472 \\ \hline
$M_{T2}$   & 0.0133 & 0.0231 \\ \hline
$\Delta R_{lb}$   & 0.0296 & 0.0726 \\ \hline
$\Delta \phi_{ll-MET}$  & 0.0100 & 0.0242 \\ \hline
$M_{lb}$  & 0.0289  & 0.0535 \\ \hline
Centrality   & 0.0074 & 0.0342 \\ \hline
$M_W$   & 0.0132 & 0.0241 \\ \hline
\end{tabular}
\caption{Separation power of features in the $e-\mu$ channel}
\label{tab:separation_power e-mu}
\end{table}

\begin{table}[h!]
\centering
\begin{tabular}{|l|c|c|}
\hline
\textbf{Feature} & \multicolumn{2}{c|}{\textbf{Separation Power}} \\
\cline{2-3}
 & \textbf{1b} & \textbf{2b} \\
\hline
$H_T$           & 0.0535 & 0.0211 \\ \hline
$\Delta R_{tW}$       & 0.0039 & 0.0035 \\ \hline
$M_{top}$       & 0.0020 & 0.0339 \\ \hline
centrality   & 0.0105 & 0.0045 \\ \hline
aplanarity   & 0.0126 & 0.0170 \\ \hline
$M_W$         & 0.0185 & 0.0110 \\ \hline
$p_T$ of $3^{rd}$ jet   & 0.0355 & 0.0150 \\ 
\hline
\end{tabular}
\caption{Separation power of features in Hadronic resolved channel}
\label{tab:separation_power hadronic resolved}
\end{table}

\begin{table}[h!]
\centering
\begin{tabular}{|l|c|c|}
\hline
\textbf{Feature} & \multicolumn{2}{c|}{\textbf{Separation Power}} \\
\cline{2-3}
 & \textbf{1b} & \textbf{2b} \\
\hline
$H_T$             & 0.0411 & 0.0292 \\ \hline
$\tau_{21}$    & 0.0183 & 0.0227 \\ \hline
softdrop mass & 0.0392 & 0.0373 \\ \hline
centrality     & 0.0170 & 0.0184 \\ \hline
aplanarity     & 0.0207 & 0.0210 \\ \hline
$M_{\mathrm{had}}$ & 0.0401 & 0.0187 \\
\hline
\end{tabular}
\caption{Separation power of features in Hadronic boosted channel}
\label{tab:separation_pwer hadronic boosted}
\end{table}

\subsection{Setup of ML Models}
\subsubsection{Boosted Decision Trees (BDT)} \label{BDT model}
The BDT model is implemented using the XGBoost framework. Running the model for different number of estimators and maximum depth for the dataset, an optimum point was found with 50 estimators and maximum depth of 2 in each tree. Further increasing depth or estimators leads to over-fitting or drop in AUC. The binary logistic loss function is used which enables the use probabilities for evaluation of model performance. 
$$-[y\log{p}+(1-y)\log({1-p})]$$ Luminosity $36000$ $pb^{-1}$ is taken for the $\sqrt{s} =13$ TeV simulated data. 
$$weight = \frac{\sigma\times W_{gen}\times\mathcal{L} }{\Sigma W_{gen}}$$ 
where
\begin{itemize}
\item$\sigma$ : cross-section
\item $W_{gen}$: generator weight
\item $\mathcal{L}$: Luminosity $=36$ $fb^{-1}$
\end{itemize}
The negative-weight events are scarce and absolute value of these weights taken for training. Due to very small absolute value, the weights are normalized and then used for training. The event weights get multiplied to the loss gradient for each iteration and each event for physically meaningful training. 
0.05 learning rate is chosen. To reduce over-training 70\%  of events are used per tree and to de-correlate trees 70\% fraction of features used per tree. 
The region in which an event falls is treated as an additional feature to aid the classification process for physics information and also to utilize test datasets correctly.
\subsubsection{Multi-Layer Perceptron (MLP)}
The Multi-Layer Perceptron (MLP) model is implemented using the PyTorch framework. The network consists of four fully connected hidden layers containing  128, 128, 64 and 32 neurons, respectively, in addition to the input and output layers. The ReLU activation function is used in the hidden layers.

The input features are scaled using the \emph{Robust Scaler} from the \emph{sklearn.preprocessing} library to reduce the effect of differences in feature magnitudes and outliers. The quantile range for the scaler is set to 10-90\%.

Binary Cross-Entropy (BCE) loss is used as the classification loss, together with the \emph{Adam} optimizer for each event, the loss contribution is weighted by the corresponding event weight to account for the physical event yields.

L2 regularization is applied to penalize, excessively large network weights, with a weight decay of $10^{-3}$. The learning rate is set to $0.005$, and the model is trained for 200 iterations. Increasing the number of iterations beyond 200 does not result in significant improvement in the model performance.

As in the BDT model, the event category (analysis region) is included as an additional input feature during training. This allows the classifier to incorporate information about the different signal and background-enriched regions.
\subsubsection{Variational Quantum Classifier (VQC)} \label{subsec: vqc methodology}

The Variational Quantum Classifier (VQC) is implemented using the \emph{Pennylane} framework with the \emph{PyTorch} interface. the input features are scaled using the \emph{Robust Scaler}, with the transformed features mapped to the interval $[0,\pi]$. The quantile range of the scaler is set to 10-90\%.

Each input feature is encoded into a corresponding qubit using parameterized $R_Y$ and $R_Z$ rotation gates. The analysis region is included as an additional input feature and is therefore encoded into an additional qubit. The quantum circuit is constructed using the \emph{StronglyEntanglingLayers} ansatz by \emph{Pennylane}, which combines parameterized single-qubit rotations with the entangling operations implemented using CNOT gates.

For the$e-\mu$ and hadronic resolved channels, 10 layers of alternating rotation and entangling operations are used. The expectation value of the Pauli-Z operator is measured from the final quantum state and used to obtain the classifier output. The output is converted to a probability for the evaluation of the classification performance.

The circuit parameters are optimized using the \emph{Adam} optimizer with a learning rate  of 0.005. The model is trained for 200 iterations. L2 regularization is applied through weight decay, with a value of $5\times 10^{-4}$ for the $e-\mu$ and hadronic resolved channels. For the hadronic boosted channel, the weight decay is reduced to $10^{-4}$ because of its smaller available statistics.

The smaller weight decay used for the VQC is motivated by the relatively small number of trainable circuit parameters and the expressive nature of the variational quantum circuit in $2^n$-dimensional Hilbert space. The implementation of the VQC using the Pennylane framework follows the approach described in references \cite{60,61}.

All three classifiers are evaluated using the same $70:30$ training-to-test data split to ensure a consistent comparison of their classification performance.
\section{Results and Discussion} \label{sec:results}
This section evaluates the three trained ML models: BDT, MLP and VQC in each of the regime of analysis: $e-mu$ channel, boosted and resolved regime of hadronic channel. For each regime of analysis the global ROC curve (True Positive Rate vs False Positive Rate) and Signal Efficiency vs Background Rejection curves for the three models are compared; the AUC (Area Under ROC Curve) and Separation Power for the three models with bootstrap uncertainties are compared; the Significance vs Signal Efficiency Curve and its characteristics: Maximum Significance, Signal efficiency at Maximum Significance, Significance gain which is the ratio of significance after applying ML to significance before applying ML, 95\% width which is the width of significance vs signal efficiency curve at 95\% of the maximum significance, and Integrated Significance which is the area under Significance vs Signal efficiency Curve for the three models are compared and the implications of this comparison are discussed in detail.
\subsection{\texorpdfstring{$e-\mu$ channel}{e-mu channel}}

\begin{table}[h]
\centering

\renewcommand{\arraystretch}{1.2}
\begin{tabular}{|c|c|c|c|c|}
\hline
\textbf{Region} & \textbf{Metric} & \textbf{BDT} & \textbf{MLP} & \textbf{VQC} \\
\hline

\multirow{2}{*}{2j1b}
& AUC &
$0.6133 \pm 0.0099$ &
$0.6183 \pm 0.0096$ &
$0.6107 \pm 0.0107$ \\
\cline{2-5}

& Separation Power &
$0.0697 \pm 0.0078$ &
$0.0651 \pm 0.0078$ &
$0.0675 \pm 0.0082$ \\
\hline

\multirow{2}{*}{2j2b}
& AUC &
$0.6852 \pm 0.0147$ &
$0.6819 \pm 0.0145$ &
$0.6898 \pm 0.0138$ \\
\cline{2-5}

& Separation Power &
$0.1432 \pm 0.0168$ &
$0.1359 \pm 0.0169$ &
$0.1594 \pm 0.0159$ \\
\hline

\end{tabular}
\caption{Region-wise classifier performance for all three models BDT, MLP and VQC in the $e-\mu$ channel.}
\label{tab:lep_auc_sep}
\end{table}

\begin{figure}[h]
\centering
\resizebox{\textwidth}{!}{
\includegraphics{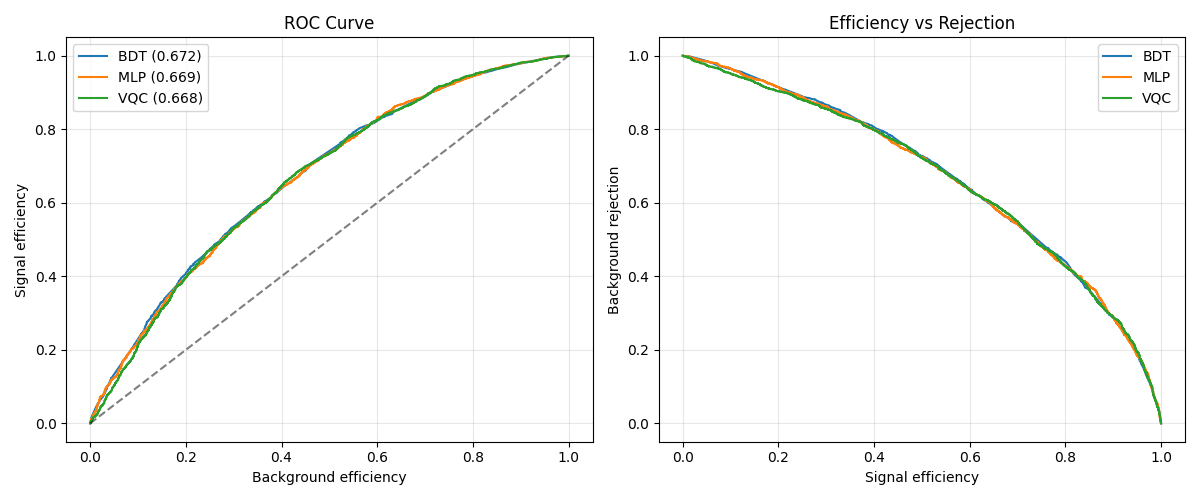}}
\caption{ROC and Signal Efficiency vs Background Rejection curve for the Global ML models BDT (blue), MLP (orange), VQC (green) for $e-\mu$ channel}
\label{fig:roc_e_mu}
\end{figure}
Table \ref{tab:lep_auc_sep} displays the AUC and separation power of the three classifiers in the signal (2j1b) region and the background (2j2b) region.  The AUC and separation powers for all the three classifiers in both the regions is similar under statistical uncertainties. Further, in figure \ref{fig:roc_e_mu}  the global ROC (Signal Efficiency vs Background Efficiency) curve and Signal Efficiency vs Background Rejection curves for all the three models nearly overlap. These curves do not represent region-wise classifier behaviour but global evaluation of the classifiers. Since the classifiers were trained with region as a feature, obtaining global classification behaviour was also possible. The AUC values suggest a moderate separation. In the signal (2j1b) region the AUC values lie in the range 0.61-0.62 and separation power in the range 0.06-0.07 while in the background (2j2b) region the AUC values lie in the range 0.68-0.69 and separation power in the range 0.13-0.16. This shows better separation in the background region than the signal region but it is largely due to fewer signal events and less overlap in variable values in the background region. This fact becomes evident when the significance metrics are analyzed for the two regions.

\begin{table*}[h]
\centering
\renewcommand{\arraystretch}{1.2}
\begin{tabular}{|c|c|c|c|c|}
\hline
\textbf{Region} &
\textbf{Metric} &
\textbf{BDT} &
\textbf{MLP} &
\textbf{VQC} \\
\hline

\multirow{5}{*}{2j1b}
& $Z_{\max}$ &
$2.0588\pm0.0524$ &
$2.0940\pm0.0532$ &
$2.0675\pm0.0492$ \\
\cline{2-5}

& $G_Z$ &
$1.0039\pm0.0045$ &
$1.0183\pm0.0096$ &
$1.0095\pm0.0038$ \\
\cline{2-5}

& $\epsilon_S^*$ &
$0.9651\pm0.0415$ &
$0.9187\pm0.0648$ &
$0.9765\pm0.0306$ \\
\cline{2-5}

& 95\% Width &
$0.2628\pm0.0258$ &
$0.2414\pm0.0530$ &
$0.2424\pm0.0445$ \\
\cline{2-5}

& Integrated Significance &
$1.6432\pm0.0528$ &
$1.6455\pm0.0498$ &
$1.6191\pm0.0512$ \\
\hline

\multirow{5}{*}{2j2b}
& $Z_{\max}$ &
$0.8378\pm0.1158$ &
$0.8458\pm0.1499$ &
$0.8462\pm0.0680$ \\
\cline{2-5}

& $G_Z$ &
$1.1471\pm0.1529$ &
$1.1664\pm0.2044$ &
$1.1584\pm0.0815$ \\
\cline{2-5}

& $\epsilon_S^*$ &
$0.5351\pm0.2099$ &
$0.6146\pm0.2991$ &
$0.5577\pm0.1597$ \\
\cline{2-5}

& 95\% Width &
$0.3484\pm0.2018$ &
$0.3568\pm0.2237$ &
$0.2983\pm0.1674$ \\
\cline{2-5}

& Integrated Significance &
$0.6885\pm0.0423$ &
$0.6728\pm0.0399$ &
$0.6911\pm0.0387$ \\
\hline

\end{tabular}
\caption{Significance metrics in the $e-\mu$ channel for all three models BDT, MLP and VQC. The rows represent the maximum significance $Z_{\max}$, significance gain $G_Z$, optimal signal efficiency $\epsilon_S^*$, 95\% significance width, and integrated significance.}
\label{tab:lep_significance}
\end{table*}

\begin{figure}[h]
\centering

\begin{subfigure}{0.48\textwidth}
\centering
\includegraphics[width=\linewidth]{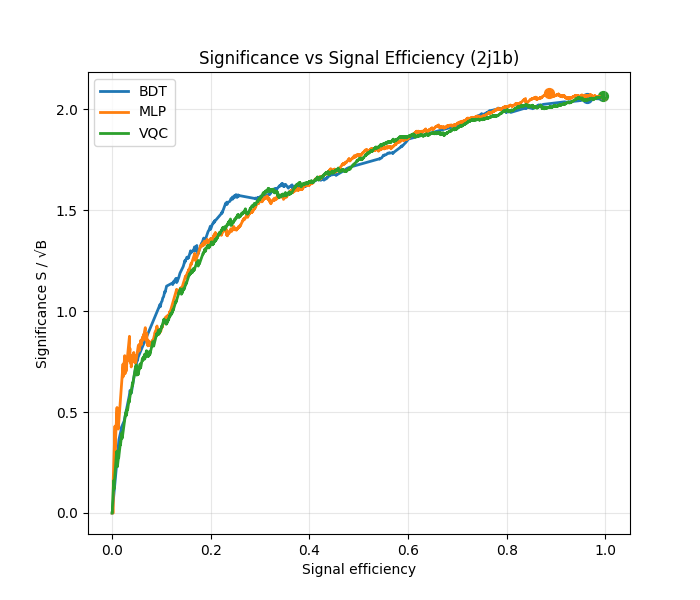}
\caption{2j1b}
\end{subfigure}
\hfill
\begin{subfigure}{0.48\textwidth}
\centering
\includegraphics[width=\linewidth]{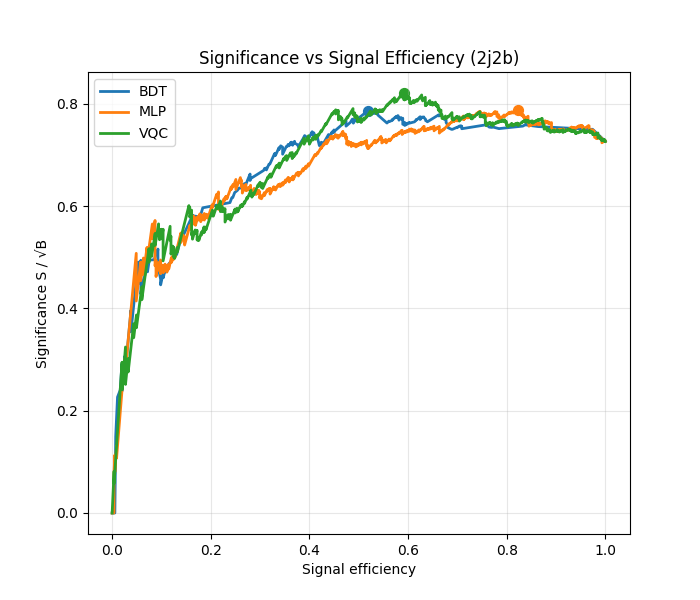}
\caption{2j2b}
\end{subfigure}

\caption{Significance vs signal efficiency curves for the trained models BDT (blue), MLP(orange) and VQC (green) in the 2j1b (left) and 2j2b (right) regions of $e-\mu$ channel .}
\label{fig:sig_e_mu}
\end{figure}

Figure \ref{fig:sig_e_mu} displays significance vs signal efficiency curves in the signal (2j1b) region and background (2j2b) region. The characteristics of these curves are calculated with bootstrap uncertainties and reported in table \ref{tab:lep_significance}. Again, all the metrics in both the regions for all the three classifiers are statistically consistent within the uncertainties. In the 2j1b region, maximum significance and integrated significance have significantly higher value than in 2j2b region, representing more signal extraction in 2j1b region. The significance gain has higher values in 2j2b region representing better separation than 2j1b region. The optimal signal efficiency values are greater than 0.9 for all the classifiers  hence \textgreater 90\% signal statistics is retained for the optimal signal efficiency in the 2j1b region. While optimal signal efficiency is near 50-60\% in the 2j2b region, representing 40-50\% signal events are not accounted for at the optimal signal efficiency point. Also, 2j2b region has highly unstable optimal signal efficiency. The 95\% width, though greater in 2j2b region, is highly unstable. This suggests that classifiers can locate a stable working region around point of maximum significance for analysis in the region 2j1b but not profoundly so in region 2j2b.
\begin{table}[h]
\centering

\renewcommand{\arraystretch}{1.2}
\begin{tabular}{|c|c|c|c|}
\hline
\textbf{Model} &
\textbf{Train AUC} &
\textbf{Test AUC} &
$\mathbf{\Delta_\sigma}$ \\
\hline

BDT &
$0.6815\pm0.0041$ &
$0.6726\pm0.0068$ &
1.13 \\
\hline

MLP &
$0.6800\pm0.0043$ &
$0.6696\pm0.0067$ &
1.27 \\
\hline

VQC &
$0.6791\pm0.0044$ &
$0.6675\pm0.0066$ &
1.40 \\
\hline

\end{tabular}
\caption{Train-test comparison of global AUC values for the three models BDT, MLP and VQC for $e-\mu$ channel.}
\label{tab:train_test_e_mu}
\end{table}
Table \ref{tab:train_test_e_mu} displays global test and train AUC for all the three models and $\frac{\Delta}{\sigma}$ for all the three models is less than 2 thus it is concluded that the models are not over-trained.
\subsection{Hadronic channel: Boosted regime}

\begin{table}[h]
\centering

\renewcommand{\arraystretch}{1.2}
\begin{tabular}{|c|c|c|c|c|}
\hline
\textbf{Region} & \textbf{Metric} & \textbf{BDT} & \textbf{MLP} & \textbf{VQC} \\
\hline

\multirow{2}{*}{1b}
& AUC &
$0.6534 \pm 0.0141$ &
$0.6687 \pm 0.0139$ &
$0.6748 \pm 0.0140$ \\
\cline{2-5}

& Separation Power &
$0.0989 \pm 0.0097$ &
$0.1346 \pm 0.0139$ &
$0.1258 \pm 0.0134$ \\
\hline

\multirow{2}{*}{2b}
& AUC &
$0.6071 \pm 0.0177$ &
$0.6012 \pm 0.0184$ &
$0.6128 \pm 0.0184$ \\
\cline{2-5}

& Separation Power &
$0.0629 \pm 0.0095$ &
$0.0739 \pm 0.0141$ &
$0.0968 \pm 0.0169$ \\
\hline

\end{tabular}
\caption{Region-wise classifier performance for all three models BDT, MLP and VQC in the hadronic boosted channel.}
\label{tab:boosted_auc_sep}
\end{table}

\begin{figure}[h]
\centering
\resizebox{\textwidth}{!}{
\includegraphics{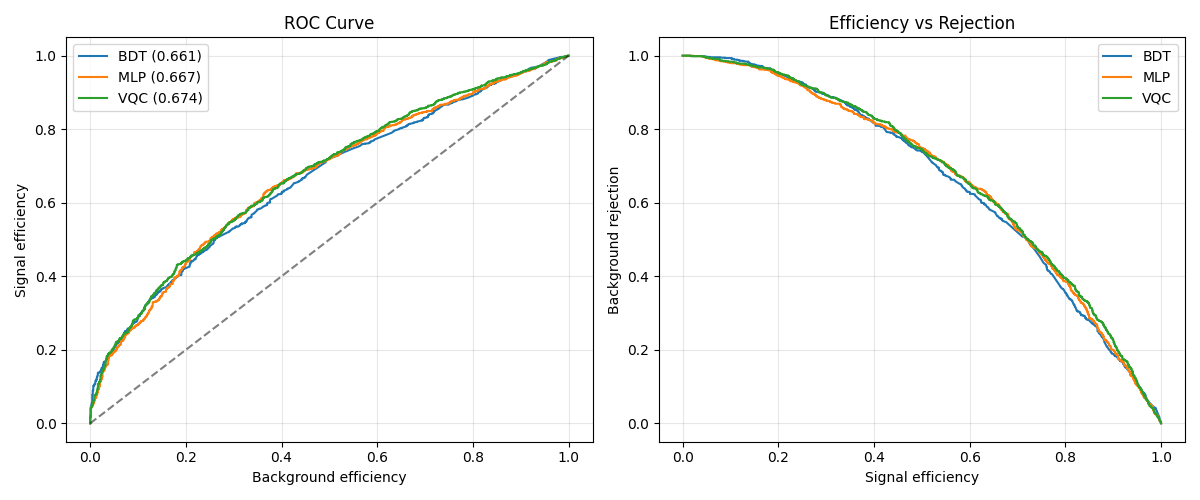}}
\caption{ROC and Signal Efficiency vs Background Rejection curve for the Global ML models BDT (blue), MLP (orange), VQC (green) for hadronic boosted channel}
\label{fig:roc_boost}
\end{figure}
Table \ref{tab:boosted_auc_sep} displays the ML metrics: AUC and separation power for the three trained models in both signal (1b) and background (2b) region. The results of this comparison become interesting for this regime. In both the signal region and background region, both the neural network models MLP and VQC have larger values of AUC and separation power as compared to BDT. In the 1b region, difference between AUC values for BDT and MLP is $0.78 \sigma$ while between BDT and VQC is $1.08 \sigma$. This is not statistically significant but an interesting result. The separation power difference between BDT and MLP is $2.04 \sigma$ and between BDT and VQC is $1.56 \sigma$. In the 2b region, the AUC values for the three models are almost equivalent under statistical uncertainties but VQC exhibits better separation power among the three in 2b region. Difference between separation power of BDT and VQC is $1.70 \sigma$ and even between MLP and VQC is $1.00 \sigma$. Figure \ref{fig:roc_boost} depicts the ROC and signal efficiency vs background rejection curves and apparently the curves for MLP and VQC are noticeably higher or better performing compared to BDT.

\begin{table*}[h]
\centering

\renewcommand{\arraystretch}{1.2}
\setlength{\tabcolsep}{4pt}
\small
\begin{tabular}{|c|c|c|c|c|}
\hline
\textbf{Region} &
\textbf{Metric} &
\textbf{BDT} &
\textbf{MLP} &
\textbf{VQC} \\
\hline

\multirow{5}{*}{1b}
& $Z_{\max}$ &
$5.879 \pm 1.689$ &
$5.013 \pm 1.357$ &
$5.585 \pm 1.535$ \\
\cline{2-5}

& $G_Z$ &
$1.838 \pm 0.516$ &
$1.568 \pm 0.422$ &
$1.742 \pm 0.473$ \\
\cline{2-5}

& $\epsilon_S^*$ &
$0.139 \pm 0.038$ &
$0.160 \pm 0.187$ &
$0.135 \pm 0.112$ \\
\cline{2-5}

& 95\% Width &
$0.023 \pm 0.028$ &
$0.098 \pm 0.230$ &
$0.047 \pm 0.144$ \\
\cline{2-5}

& Integrated Significance &
$3.126 \pm 0.198$ &
$2.978 \pm 0.156$ &
$3.158 \pm 0.194$ \\
\hline

\multirow{5}{*}{2b}
& $Z_{\max}$ &
$2.434 \pm 0.624$ &
$1.748 \pm 0.114$ &
$1.852 \pm 0.322$ \\
\cline{2-5}

& $G_Z$ &
$1.428 \pm 0.359$ &
$1.020 \pm 0.052$ &
$1.090 \pm 0.189$ \\
\cline{2-5}

& $\epsilon_S^*$ &
$0.216 \pm 0.296$ &
$0.911 \pm 0.224$ &
$0.687 \pm 0.406$ \\
\cline{2-5}

& 95\% Width &
$0.120 \pm 0.279$ &
$0.241 \pm 0.196$ &
$0.290 \pm 0.287$ \\
\cline{2-5}

& Integrated Significance &
$1.399 \pm 0.087$ &
$1.377 \pm 0.081$ &
$1.414 \pm 0.086$ \\
\hline

\end{tabular}
\caption{Significance-based performance metrics in the hadronic boosted channel for the BDT, MLP and VQC classifiers. The rows correspond to the maximum significance $Z_{\max}$, significance gain $G_Z$, optimal signal efficiency $\epsilon_S^*$, 95\% significance width, and integrated significance.}
\label{tab:boosted_significance}
\end{table*}

\begin{figure}[h]
\centering
\begin{subfigure}{0.48\textwidth}
\centering
\includegraphics[width=\linewidth]{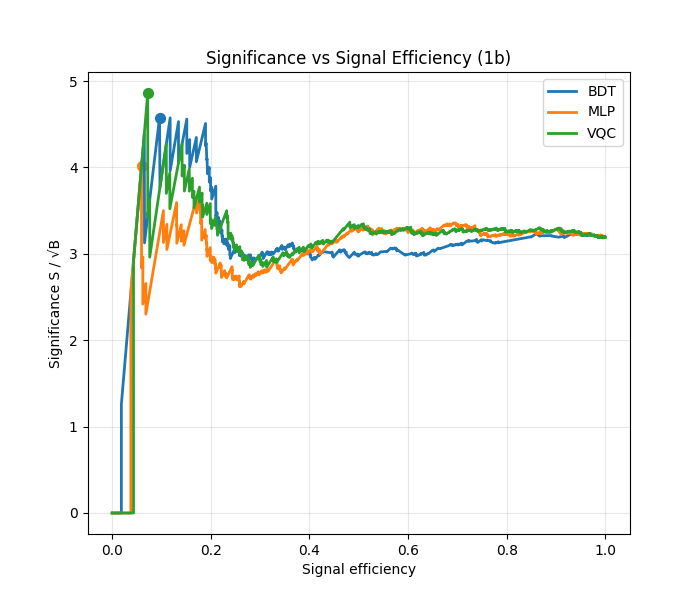}
\caption{1b}
\end{subfigure}
\hfill
\begin{subfigure}{0.48\textwidth}
\centering
\includegraphics[width=\linewidth]{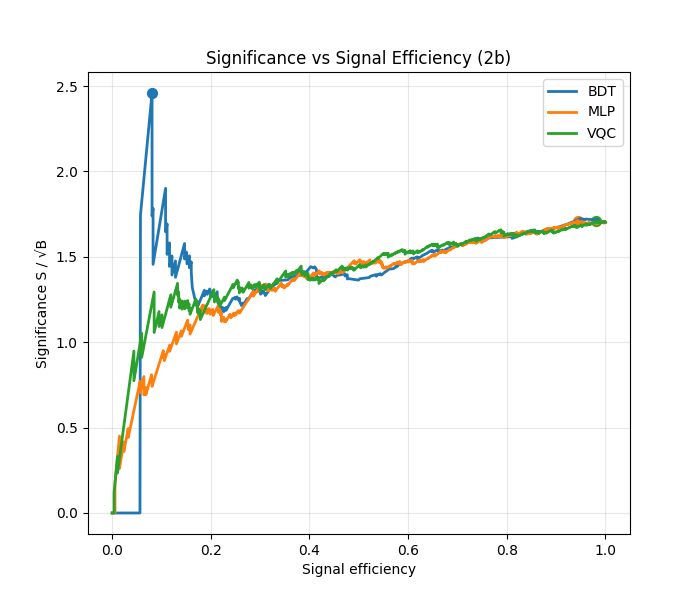}
\caption{2b}
\end{subfigure}

\caption{Significance vs signal efficiency curves for the trained models BDT (blue), MLP(orange) and VQC (green) in the 2j1b (left) and 2j2b (right) regions of hadronic boosted channel .}
\label{fig:sig_boost}
\end{figure}
Figure \ref{fig:sig_boost} depicts the significance vs signal efficiency curves for 1b and 2b regions. It is observed from the plots that they exhibit very sharp peaks at low signal efficiencies. The large uncertainties in the maximum significance and 95\% width values in both the regions in table \ref{tab:boosted_significance} are evidence for the unstable nature of the peaks. In the 2b region, the BDT attains a very high maximum significance and significance gain values but the peak is very unstable (values have high uncertainties) and hence its superiority can not be claimed with statistical confidence. At higher signal efficiencies, the curves for all the models overlap in 2b region while BDT noticeably under-performs at higher signal efficiencies in 1b region. The integrated significance values however, in both regions are statistically consistent across all the three models. Thus, in the hadronic boosted regime MLP and VQC exhibited better ML metrics while BDT exhibited higher but unstable significance regions while MLP and VQC show relatively stable significance curves. Nevertheless, integrated significance for both regions remain similar under statistical constraints in both regions.
\begin{table}[h]
\centering
\renewcommand{\arraystretch}{1.2}
\begin{tabular}{|c|c|c|c|}
\hline
\textbf{Model} &
\textbf{Train AUC} &
\textbf{Test AUC} &
$\mathbf{\Delta_\sigma}$ \\
\hline

BDT &
$0.6623 \pm 0.0061$ &
$0.6612 \pm 0.0095$ &
0.10 \\
\hline

MLP &
$0.6711 \pm 0.0067$ &
$0.6666 \pm 0.0101$ &
0.38 \\
\hline

VQC &
$0.6764 \pm 0.0062$ &
$0.6740 \pm 0.0096$ &
0.21 \\
\hline

\end{tabular}
\caption{Train-test comparison of global AUC values for the three models BDT, MLP and VQC in the hadronic boosted channel.}
\label{tab:boosted_train_test}
\end{table}

Table \ref{tab:boosted_train_test} displays global train and test AUCs of all the three models and clearly the models trained are not overtrained.
\subsection{Hadronic channel: Resolved regime}

\begin{table}[h]
\centering

\renewcommand{\arraystretch}{1.2}
\begin{tabular}{|c|c|c|c|c|}
\hline
\textbf{Region} & \textbf{Metric} & \textbf{BDT} & \textbf{MLP} & \textbf{VQC} \\
\hline

\multirow{2}{*}{1b}
& AUC &
$0.6469 \pm 0.0062$ &
$0.6621 \pm 0.0062$ &
$0.6582 \pm 0.0062$ \\
\cline{2-5}

& Separation Power &
$0.0777 \pm 0.0059$ &
$0.0928 \pm 0.0064$ &
$0.0858 \pm 0.0068$ \\
\hline

\multirow{2}{*}{2b}
& AUC &
$0.6414 \pm 0.0088$ &
$0.6261 \pm 0.0100$ &
$0.6377 \pm 0.0098$ \\
\cline{2-5}

& Separation Power &
$0.0823 \pm 0.0092$ &
$0.0616 \pm 0.0078$ &
$0.0773 \pm 0.0083$ \\
\hline

\end{tabular}
\caption{Region-wise classifier performance for the three models BDT, MLP and VQC in the hadronic resolved channel.}
\label{tab:resolved_auc_sep}
\end{table}

\begin{figure}[h]
\centering
\resizebox{\textwidth}{!}{
\includegraphics{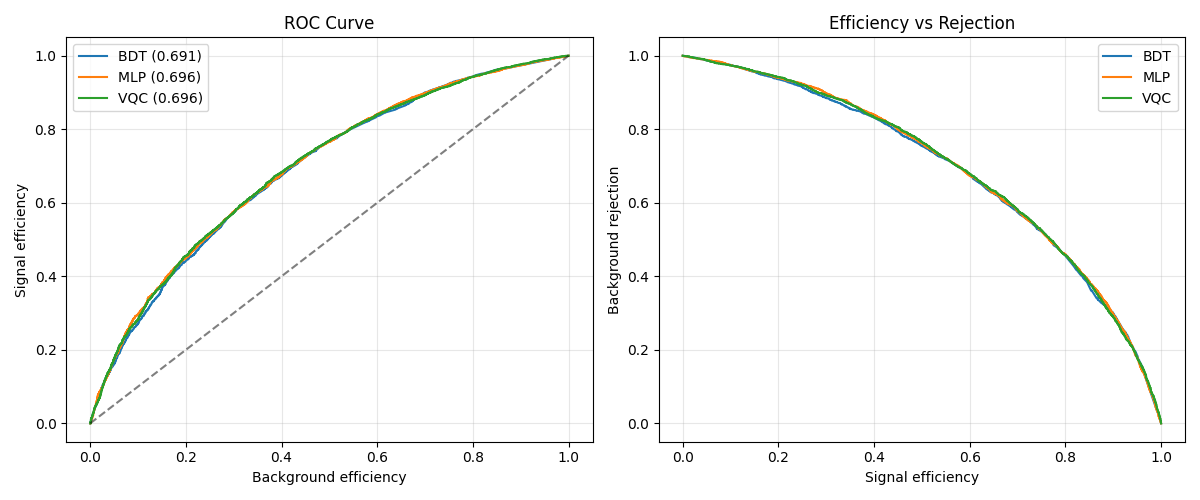}}
\caption{ROC and Signal Efficiency vs Background Rejection curve for the Global ML models BDT (blue), MLP (orange), VQC (green) for hadronic resolved channel.}
\label{fig:roc_res}
\end{figure}

Table \ref{tab:resolved_auc_sep} represents ML metrics for each of the three trained models for signal (1b) region and background (2b) region of the hadronic resolved channel. In the 1b region MLP has greater central value for AUC and separation power. The difference between AUC of BDT and MLP is $1.73 \sigma$ and between their separation power is $1.74 \sigma$, between AUC of BDT and VQC is $1.29 \sigma$ and between their separation power is $0.90 \sigma$. The difference between the ML metrics for MLP and VQC is not significant, however, MLP exhibits numerically higher central values here in the 1b region. In the 2b region, BDT has higher central values for AUC and separation power. The difference between AUC of BDT and MLP is $1.15 \sigma$ and between their separation powers is $1.72 \sigma$. However, VQC has values close to BDT and exhibits better values than MLP but still the difference is statistically insignificant. In figure \ref{fig:roc_res} the global ROC and signal efficiency vs background rejection curve for all the three models almost overlap.

\begin{table*}[h]
\centering

\renewcommand{\arraystretch}{1.2}
\setlength{\tabcolsep}{4pt}
\small
\begin{tabular}{|c|c|c|c|c|}
\hline
\textbf{Region} &
\textbf{Metric} &
\textbf{BDT} &
\textbf{MLP} &
\textbf{VQC} \\
\hline

\multirow{5}{*}{1b}
& $Z_{\max}$ &
$8.858 \pm 0.137$ &
$8.971 \pm 0.146$ &
$8.909 \pm 0.149$ \\
\cline{2-5}

& $G_Z$ &
$1.030 \pm 0.007$ &
$1.044 \pm 0.009$ &
$1.037 \pm 0.009$ \\
\cline{2-5}

& $\epsilon_S^*$ &
$0.893 \pm 0.024$ &
$0.864 \pm 0.022$ &
$0.856 \pm 0.044$ \\
\cline{2-5}

& 95\% Width &
$0.351 \pm 0.041$ &
$0.393 \pm 0.055$ &
$0.406 \pm 0.050$ \\
\cline{2-5}

& Integrated Significance &
$7.282 \pm 0.146$ &
$7.478 \pm 0.153$ &
$7.413 \pm 0.154$ \\
\hline

\multirow{5}{*}{2b}
& $Z_{\max}$ &
$2.516 \pm 0.071$ &
$2.495 \pm 0.060$ &
$2.533 \pm 0.060$ \\
\cline{2-5}

& $G_Z$ &
$1.019 \pm 0.013$ &
$1.007 \pm 0.006$ &
$1.024 \pm 0.010$ \\
\cline{2-5}

& $\epsilon_S^*$ &
$0.813 \pm 0.096$ &
$0.937 \pm 0.077$ &
$0.884 \pm 0.064$ \\
\cline{2-5}

& 95\% Width &
$0.448 \pm 0.058$ &
$0.387 \pm 0.058$ &
$0.358 \pm 0.067$ \\
\cline{2-5}

& Integrated Significance &
$2.060 \pm 0.064$ &
$1.989 \pm 0.054$ &
$2.029 \pm 0.061$ \\
\hline

\end{tabular}
\caption{Significance metrics in the hadronic resolved channel for the BDT, MLP and VQC classifiers. The rows correspond to the maximum significance $Z_{\max}$, significance gain $G_Z$, optimal signal efficiency $\epsilon_S^*$, 95\% significance width, and integrated significance.}
\label{tab:resolved_significance}
\end{table*}

\begin{figure}[h]
\centering
\begin{subfigure}{0.48\textwidth}
\centering
\includegraphics[width=\linewidth]{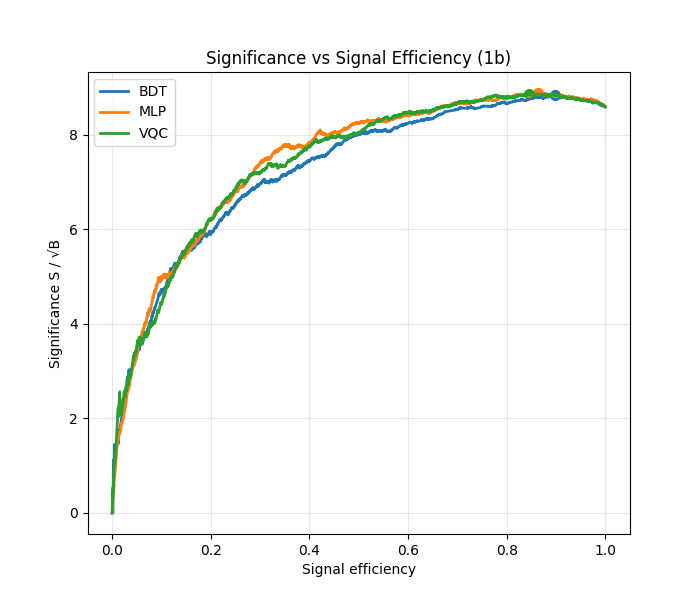}
\caption{1b}
\end{subfigure}
\hfill
\begin{subfigure}{0.48\textwidth}
\centering
\includegraphics[width=\linewidth]{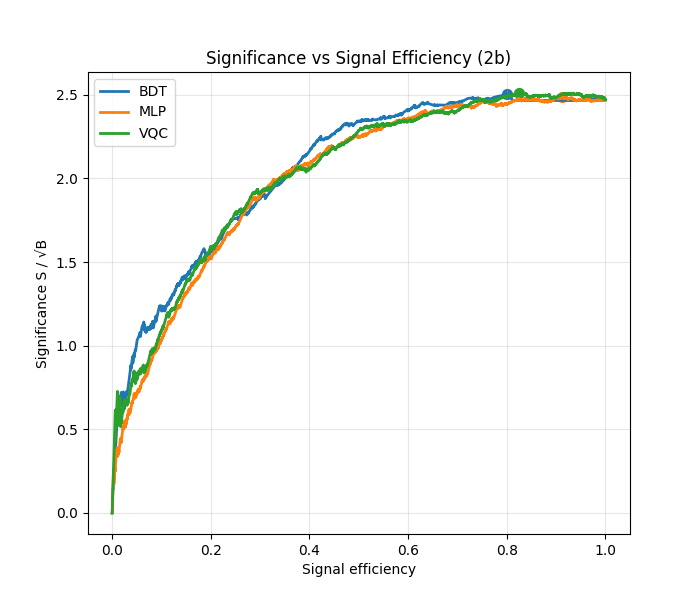}
\caption{2b}
\end{subfigure}

\caption{Significance vs signal efficiency curves for the trained models BDT (blue), MLP (orange) and VQC (green) in the 2j1b (left) and 2j2b (right) regions of hadronic resolved channel .}
\label{fig:sig_res}
\end{figure}

Figure \ref{fig:sig_res} depicts the significance vs signal efficiency curves for the signal (1b) and background (2b) regions. The BDT curve appears slightly below the other two curves in the 1b region and slightly above the two curves in the 2b region. However, when bootstrap uncertainties are taken into account, the differences almost vanish. Table \ref{tab:resolved_significance} reveals that all the significance-related metrics for all the three models in this regime are statistically very similar. Only one interesting comparison appears that the 95\% width in 2b region for BDT has a great value compared to that of VQC, there is $1.02 \sigma$ difference.
\begin{table}[h]
\centering

\renewcommand{\arraystretch}{1.2}
\begin{tabular}{|c|c|c|c|}
\hline
\textbf{Model} &
\textbf{Train AUC} &
\textbf{Test AUC} &
$\mathbf{\Delta_\sigma}$ \\
\hline

BDT &
$0.6983 \pm 0.0031$ &
$0.6911 \pm 0.0050$ &
1.23 \\
\hline

MLP &
$0.7031 \pm 0.0032$ &
$0.6969 \pm 0.0048$ &
1.10 \\
\hline

VQC &
$0.7026 \pm 0.0031$ &
$0.6957 \pm 0.0049$ &
1.20 \\
\hline

\end{tabular}
\caption{Train-test comparison of global AUC values for the three models BDT, MLP and VQC in the hadronic resolved channel.}
\label{tab:resolved_train_test}
\end{table}
Table \ref{tab:resolved_train_test} gives evidence that there is no statistically significant over-training of the three models based on the train and test AUC values in the hadronic resolved channel.

\section{Conclusion and Future Scope}
The study was aimed at evaluating the utility of Quantum Neural Networks for the problem of background discrimination in high energy physics. For this purpose, the Variational Quantum Classifier was tested in three different physics regimes for the same background discrimination problem: $tW$ vs $t \bar{t}$ and its performance compared with standard ML algorithms: BDT and MLP routinely employed for such problems. The summary of the results and comparison are presented in this section.

\vspace{3pt}

The performance of the Variational Quantum Classifier algorithm in all the three physics regimes for all the regions remained statistically similar to the other two models: BDT and MLP. For this problem with the selected features and limited statistics all models could only provide moderate discrimination (AUC range 0.60-0.70). 

\vspace{3pt}

The ML and significance metrics in the leptonic channel remain almost similar for all the three models. This result conveys the indistinguishability of the ML models in this physics channel. Hence, VQC can be successfully applied to achieve same performance as standard classifiers in this physics channel. 

\vspace{3pt}

In the hadronic channel, for both regimes: boosted and resolved, The neural network algorithms MLP and VQC displayed higher values of ML metrics: AUC and separation power for the signal region, although the differences were small and comparable to the statistical uncertainties. In the background region, however, BDT showcased higher central values for AUC and separation power in the resolved channel while VQC displayed higher central values of separation power in the boosted channel. In the boosted channel, BDT appears to extract high significance but in the form of very sharp and unstable peaks. Due to high bootstrap uncertainties in the peak value no statistically confident claim of its superiority can be made in the boosted physics channel. This observation is also supported by similar values of integrated significance for all the three models in the boosted channel. VQC and MLP appear to have more stable curves in the background region of the boosted channel. In the boosted channel, VQC offers a unique behaviour mimicking the stability of neural networks like MLP while offering a slightly higher or comparable values of significance. In the resolved channel the significance metrics were again very similar despite very small uncertainties.

\vspace{3pt}

Although model parameters were optimized for avoiding over-training, different feature-sets and sample sizes may change the results. However, analyzing the results of all physics channel in this study, it can be safely claimed that VQC can achieve performance comparable to standard ML models like BDT and MLP. Further, Neural Networks may offer a smoother significance curve in noisy physics channels like the boosted hadronic channel in this study. VQC could combine its quantum superposition and entanglement properties added to its neural network architecture to offer an improved solution. 

\vspace{3pt}

Thus, the study establishes that the Variational Quantum Classifier algorithm may be utilized for background discrimination in high energy physics by identifying that it performs comparably to other standard ML algorithms like MLP and BDT. VQC might find use in various other HEP data analysis problems where Neural Networks outperform BDT. For example, the study in reference \cite{91} perfroms a comparison between neural networks and BDT for $t\bar{t}H(b\bar{b})$ analysis and finds that neural networks perform slightly better than BDT while exhibiting no additional training bias. Studies in references \cite{87, 88, 89} and \cite{90} suggest that neural networks might offer a better model for top-tagging, jet reconstruction and particle identification. If VQC be applied to these problems or specific problems for analysis where the neural network architecture better exploits the variables, it might show better results than conventional models due to its unique properties of quantum entanglement and superposition. 

\acknowledgments

We acknowledge the support of CERN and CMS for making simulated datasets available on CERN open data portal.The datasets \cite{62,63,64,65,66,67,68,69,70,71,72,73,74,75,76,77,78,79,80,81,82} taken from CERN open data portal were utilized in this study. 

We further acknowledge the support of the Department of Physics at Hindu College and the University of Delhi for providing the necessary environment, which made this research possible. 

We also acknowledge the assistance provided by the AI tool ChatGPT in debugging the lengthy codes involved in the analysis and also acknowledge its parent company OpenAI.
 
\appendix

\section{Backgrounds of tW }
Although $tW$ has various backgrounds (like QCD jets and W+jets) and often with far larger cross-section than $t\bar{t}$, they are suppressed with appropriate Physics cuts but $t\bar{t}$ has very similar final state and can not be separated from $tW$ events simply using cuts \cite{49,50,51,52,53}.
\begin{figure}[h]
\centering

\begin{subfigure}{0.48\textwidth}
\centering
\includegraphics[width=\linewidth]{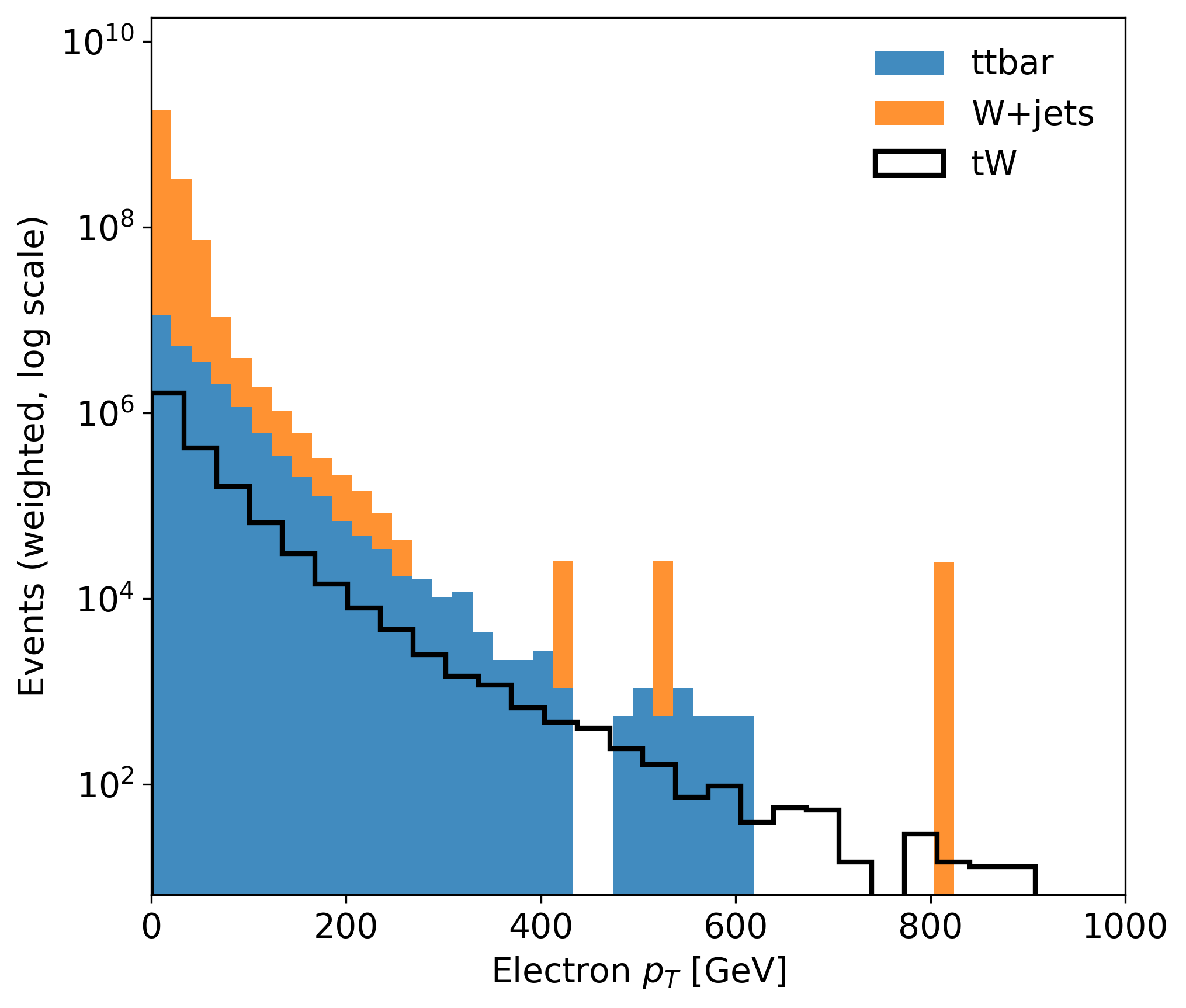}
\caption{electron $p_T$ without cuts}
\end{subfigure}
\hfill
\begin{subfigure}{0.48\textwidth}
\centering
\includegraphics[width=\linewidth]{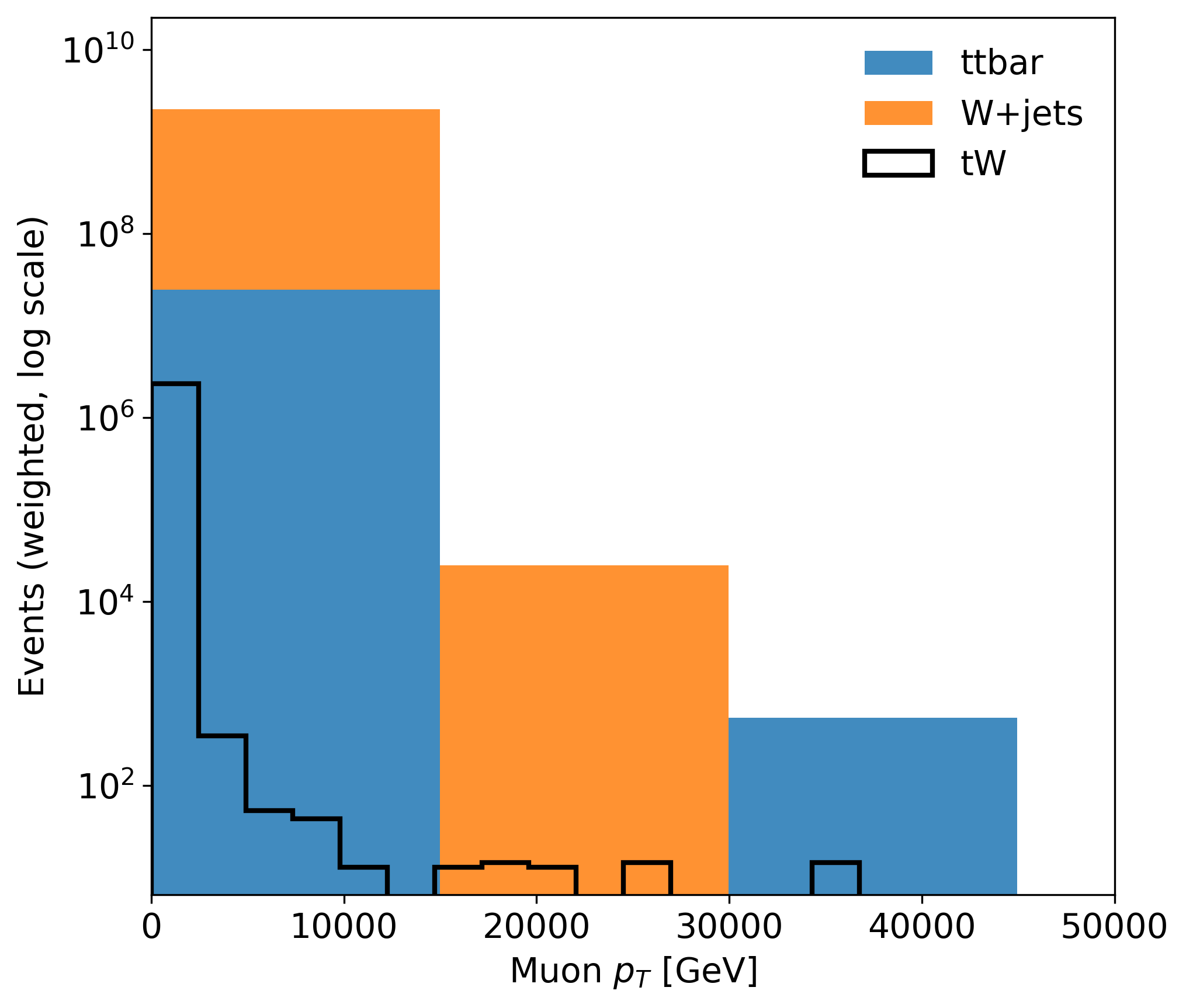}
\caption{$\mu$ $p_T$ without cuts}
\end{subfigure}

\vspace{0.3cm}

\begin{subfigure}{0.48\textwidth}
\centering
\includegraphics[width=\linewidth]{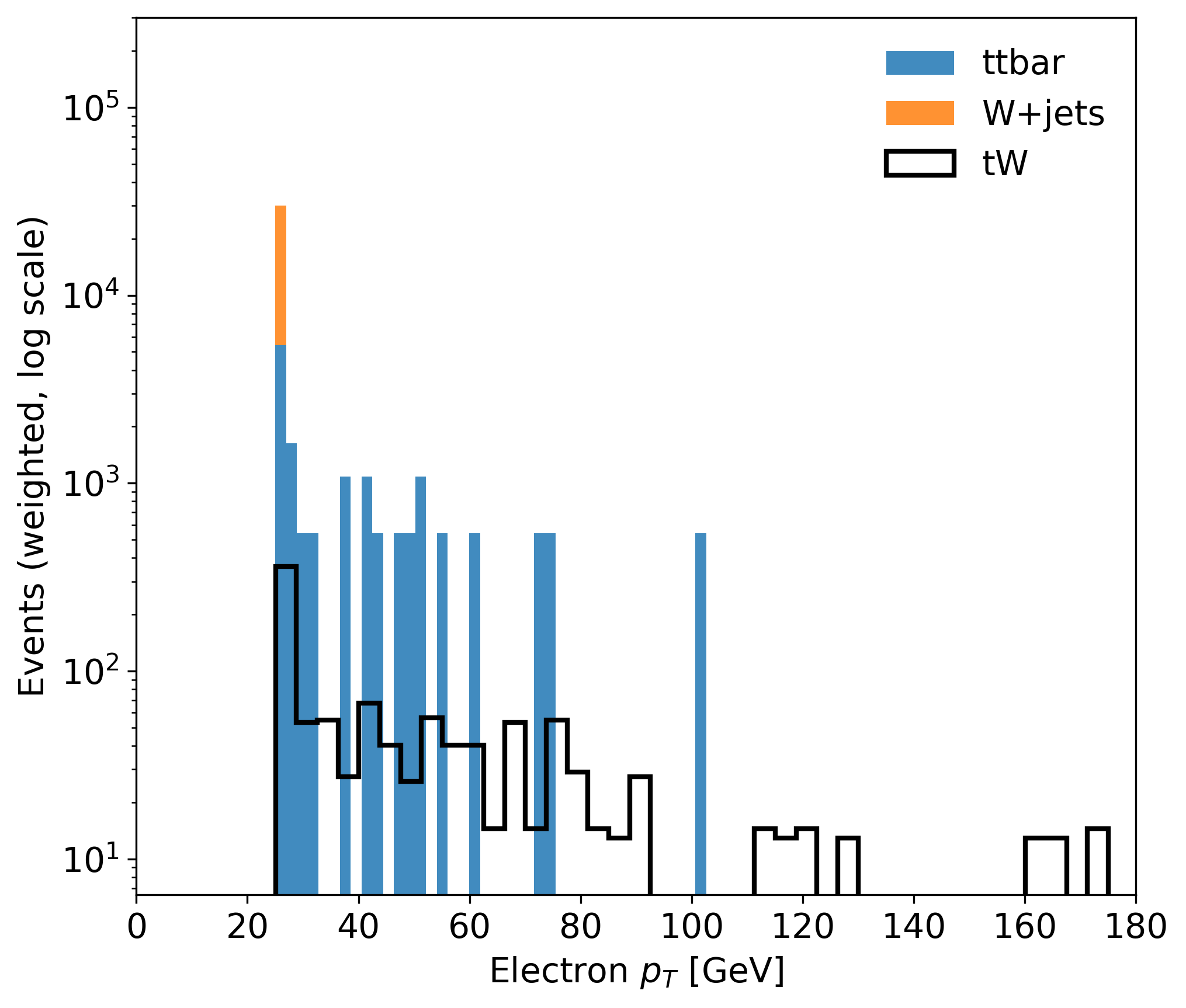}
\caption{electron $p_T$ after cuts}
\end{subfigure}
\hfill
\begin{subfigure}{0.48\textwidth}
\centering
\includegraphics[width=\linewidth]{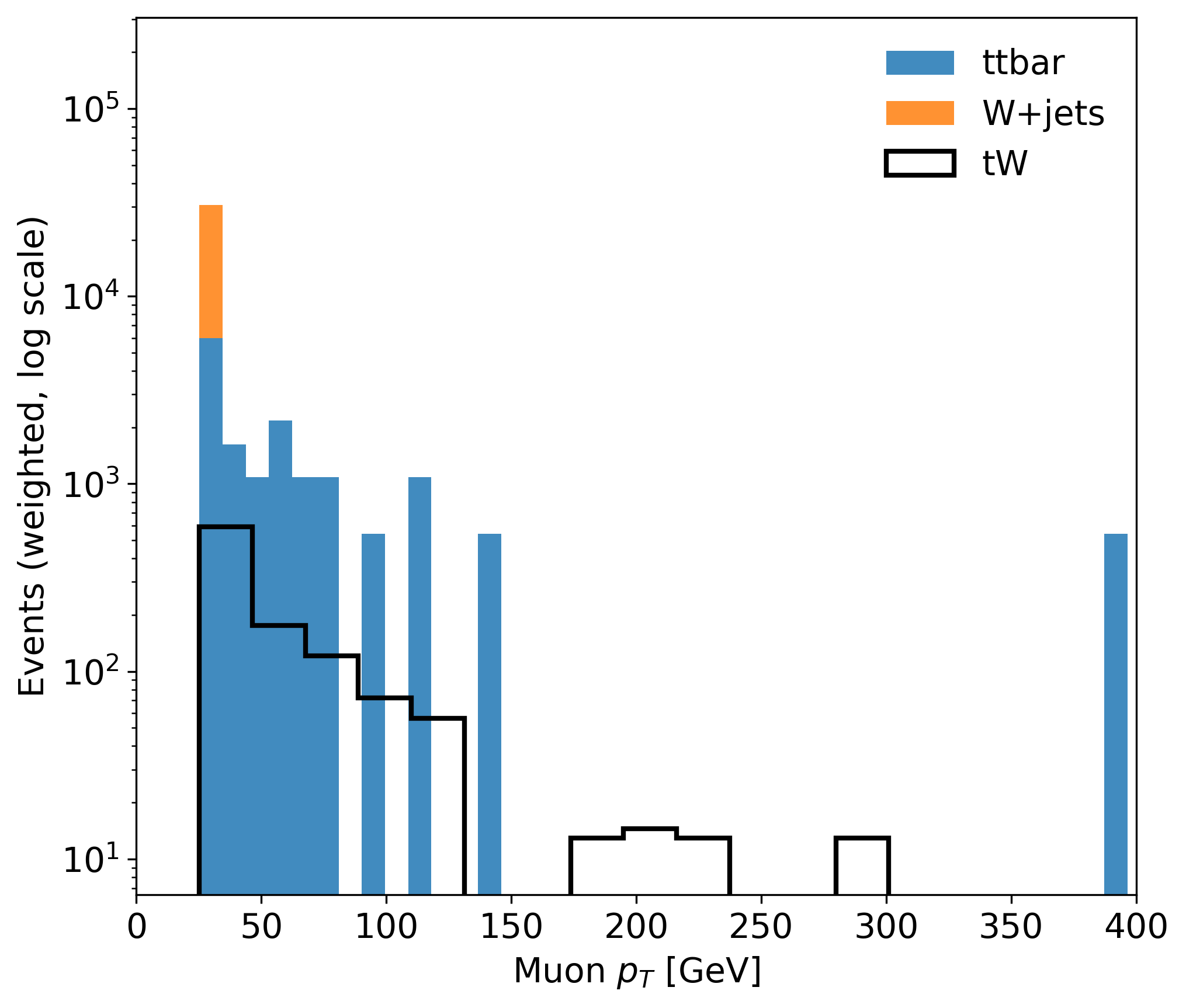}
\caption{$\mu$ $p_T$ after cuts}
\end{subfigure}

\caption{suppression of W+jets in $e-\mu$ channel (Note: the data for these plots has been taken from CERN open data portal datasets. For exact datasets and cross-sections refer table \ref{tab:datasets})}
\label{fig:e-mu1}
\end{figure}

Figure \ref{fig:e-mu1} shows how the W+jets background can be suppressed for the $e-\mu$ channel. The cuts applied for this suppression are:
\begin{itemize}[itemsep=1.5pt, topsep=3pt]
\item{$\mu-{p_T} > 25$}
\item{$e-{p_T} > 25 $}
\item{$ \mu-|\eta|< 2.4$ } 
\item{$ e-|\eta|< 2.5$ } 
\item{$\mu_{iso} < 0.15$}
\item{$e_{iso} < 0.15$}
\item{$\mu_q \times el_q < 0$}
\item{$MET > 30$}
\item{$N_j \leq 2$}
\item{$N_b = 1$}
\end{itemize}
\begin{figure}[h]
\centering

\begin{subfigure}{0.48\textwidth}
\centering
\includegraphics[width=\linewidth]{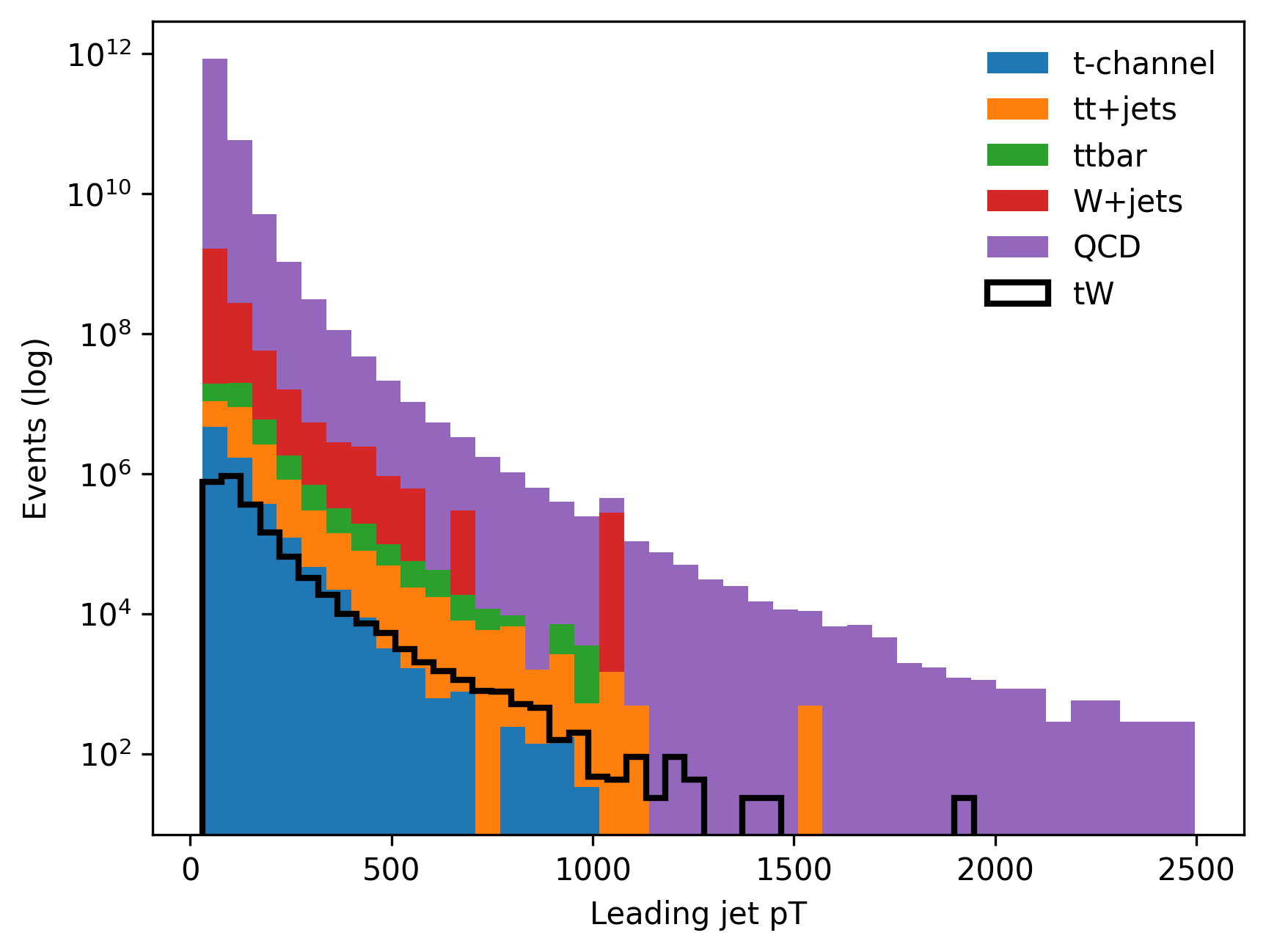}
\caption{leading jet $p_T$ without cuts}
\end{subfigure}
\hfill
\begin{subfigure}{0.48\textwidth}
\centering
\includegraphics[width=\linewidth]{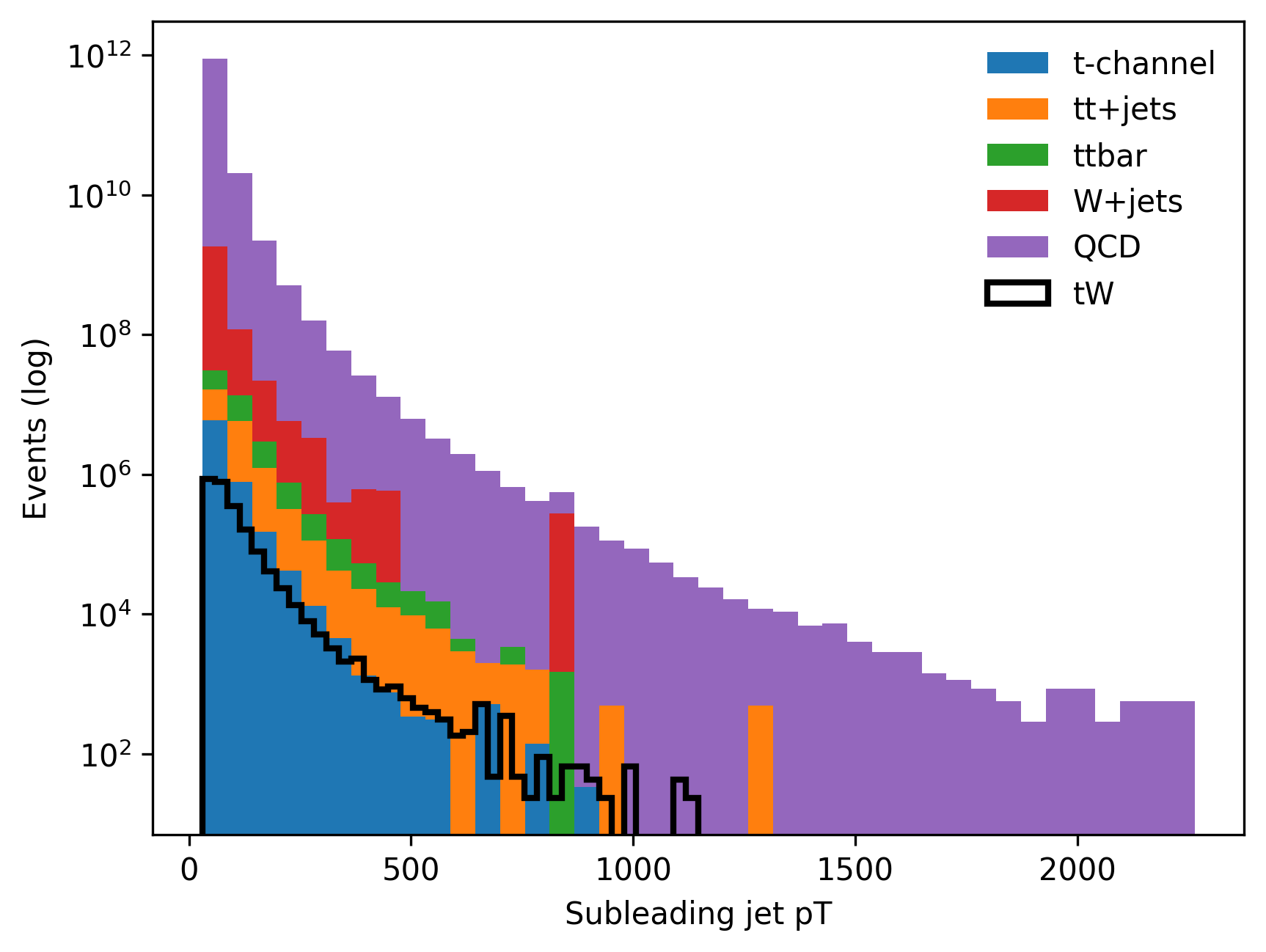}
\caption{subleading jet $p_T$ without cuts}
\end{subfigure}

\vspace{0.3cm}

\begin{subfigure}{0.48\textwidth}
\centering
\includegraphics[width=\linewidth]{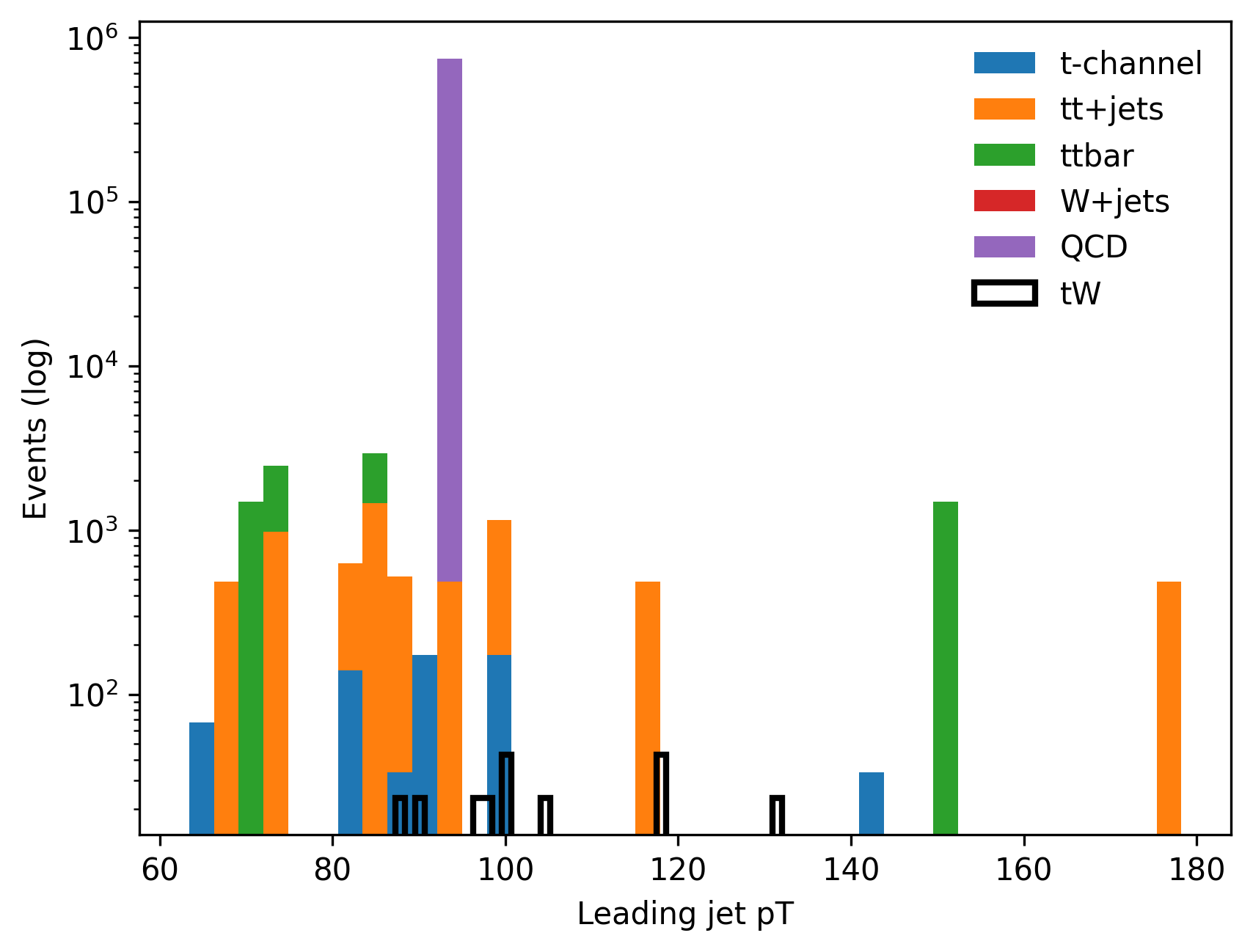}
\caption{leading jet $p_T$ after cuts}
\end{subfigure}
\hfill
\begin{subfigure}{0.48\textwidth}
\centering
\includegraphics[width=\linewidth]{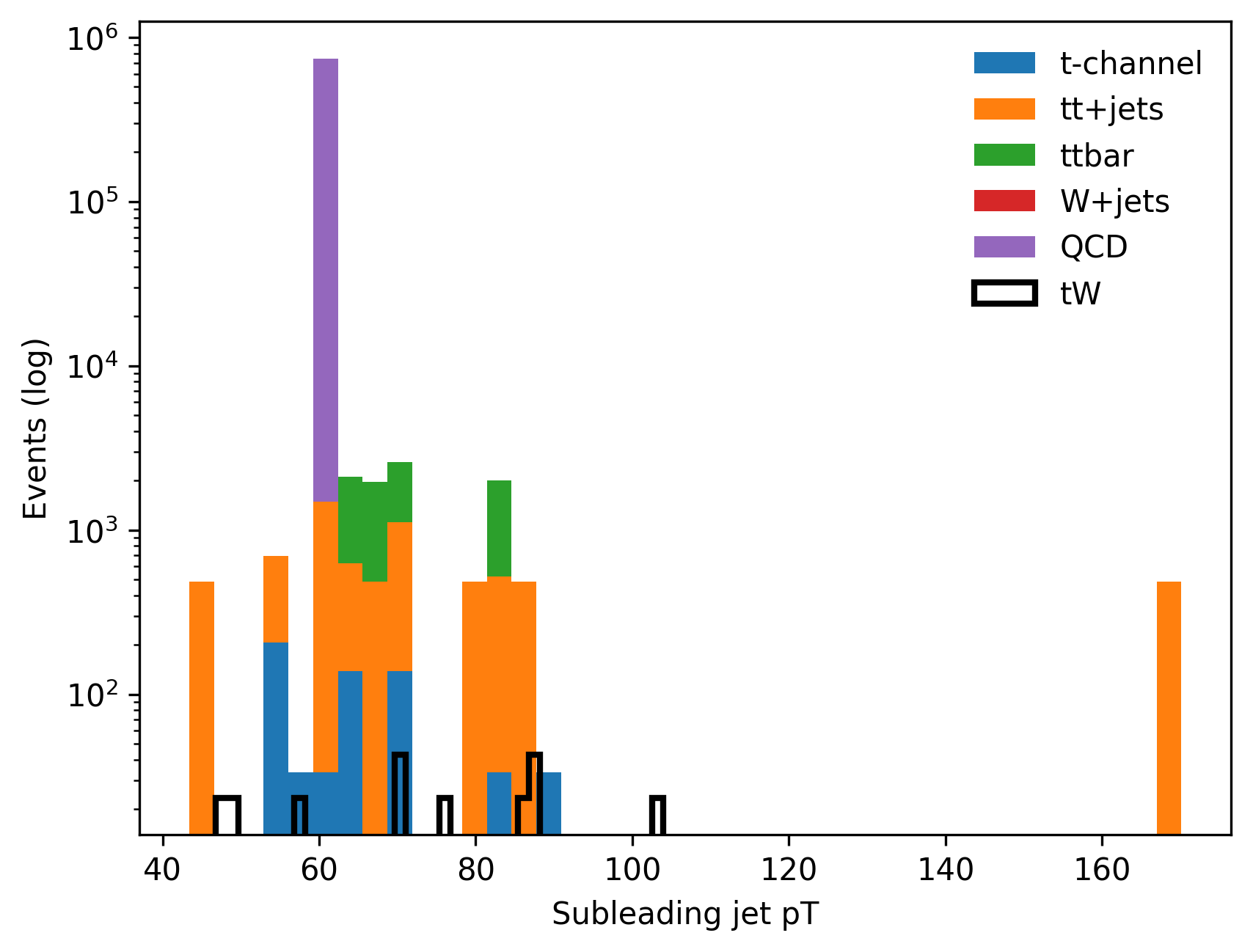}
\caption{subleading jet $p_T$ after cuts }
\end{subfigure}

\caption{suppression of QCD jets and W+jets in hadronic channel. (Note: the data for these plots has been taken from CERN open data portal datasets. For exact datasets and cross-sections refer table \ref{tab:datasets})}
\label{fig:e-mu2}
\end{figure}
The hadronic channel has backgrounds W+jets, QCD jets, t-channel single top, $t\bar{t}$ and $t\bar{t}$+jets, and upon applying appropriate cuts large cross-section W+jets, and QCD jets backgrounds can be suppressed. This is shown in figure \ref{fig:e-mu2}.The cuts applied are:
\begin{itemize}[itemsep=1.5pt, topsep=3pt]
\item{$3 \leq N_j \leq 4$}
\item{ $N_b \geq 1$} 
\item{$jet_1-{p_T} > 30$}
\item{$jet_2-{p_T} > 10$}
\item{$150<m_{top}<200$ (reconstructed top mass)}
\item{$sphericity > 0.5$} 
\item{$aplanarity > 0.04$}
\item{$w_{balance} > 0.5$ ($w_{balance} = \frac{min(jet_1-p_T,jet_2-p_T)}{max(jet_1-p_T,jet_2-p_T)}$})
\item{$\Delta\phi < 1.0$}
\end{itemize}



\bibliographystyle{JHEP}
\bibliography{biblio.bib}
\end{document}